\documentclass[twocolumn, times]{aastex63}

\usepackage{amsmath,verbatim, nicefrac,color, soul}
\usepackage{natbib}
\usepackage[caption=false]{subfig}
\newcommand\Hl[1]{\colorbox{yellow}}

\shorttitle{Simultaneous Dual-Frequency Scintillation Arc Survey}
\shortauthors{J. E. Turner \lowercase{et al}.}

\begin{document}
\title{A Simultaneous Dual-Frequency Scintillation Arc Survey of Six Bright Canonical Pulsars Using the Upgraded Giant Metrewave Radio Telescope} 

\author[0000-0002-2451-7288]{Jacob E. Turner}
\affiliation{Department of Physics and Astronomy, West Virginia University, P.O. Box 6315, Morgantown, WV 26506, USA}
\affiliation{Center for Gravitational Waves and Cosmology, West Virginia University, Chestnut Ridge Research Building, Morgantown, WV 26505, USA}
\affiliation{Green Bank Observatory, P.O. Box 2, Green Bank, WV 24944, USA}

\author[0000-0002-0863-7781]{Bhal Chandra Joshi}
\affiliation{National Centre for Radio Astrophysics, Tata Institute of Fundamental Research, Post Bag 3, Ganeshkhind, Pune - 411007, India}

\author[0000-0001-7697-7422]{Maura A. McLaughlin}
\affiliation{Department of Physics and Astronomy, West Virginia University, P.O. Box 6315, Morgantown, WV 26506, USA}
\affiliation{Center for Gravitational Waves and Cosmology, West Virginia University, Chestnut Ridge Research Building, Morgantown, WV 26505, USA}

\author[0000-0002-1797-3277]{Daniel R. Stinebring}
\affiliation{Department of Physics and Astronomy, Oberlin College, Oberlin, OH 44074, USA}


\begin{abstract}
We use the upgraded Giant Metrewave Radio Telescope to measure scintillation arc properties in six bright canonical pulsars with simultaneous dual frequency coverage. These observations at frequencies from 300 to 750 MHz allowed for detailed analysis of arc evolution across frequency and epoch. We perform more robust determinations of frequency dependence for arc curvature, scintillation bandwidth, and scintillation timescale, and comparison between arc curvature and pseudo-curvature than allowed by single-frequency-band-per-epoch measurements, which we find to agree with theory and previous literature. We find a strong correlation between arc asymmetry and arc curvature, which we have replicated using simulations, and attribute to a bias in the Hough transform approach to scintillation arc analysis. Possible evidence for an approximately week long timescale over which a given scattering screen dominates signal propagation was found by tracking visible scintillation arcs in each epoch in PSR J1136+1551. The inclusion of a 155 minute observation allowed us to resolve the scale of scintillation variations on short timescales, which we find to be directly tied to the amount of ISM sampled over the observation. Some of our pulsars showed either consistent or emerging asymmetries in arc curvature, indicating instances of refraction across their lines of sight. Significant features in various pulsars, such as multiple scintillation arcs in PSR J1136+1551 and flat arclets in PSR J1509+5531, that have been found in previous works, were also detected. The simultaneous multiple band observing capability of the upgraded GMRT shows excellent promise for future pulsar scintillation work.
\end{abstract}

\section{Introduction}
The scintillation of pulsar emission occurs as the result of its propagation through non-uniform distributions of free electrons in the ionized interstellar medium (ISM). This interaction results in frequency-dependent and time-evolving variations in the flux density of the pulsar signal as measured at a detector. When these variations are examined across observing frequency and time in so-called dynamic spectra, representations of the change in the pulsar signal's intensity across frequency and time, for a given observation, they can provide valuable insight into the structure of these electron density variations along our line of sight (LOS) to a given pulsar. Information can be gained about the ISM structure along the LOS by examining the parabolic arcs, known as scintillation arcs, that can emerge by examining the power spectrum of the dynamic spectrum, generally known as the secondary spectrum \citep{OG_arcs}. Successful analysis requires sufficient resolution in time and frequency of the scintles, or bright patches in the dynamic spectrum, due to constructive interference between different ray paths through the ISM. Refractive shifts due to, e.g., wedge-like plasma structures cause scintillation drift patterns that cause asymmetries in the scintillation arc intensity distribution and an offset of the parabola origin \citep{cordes_2006_refraction}. Some current hypotheses on the physical origins of these arcs postulate that they originate from compressed plasma along the boundaries of 50$-$100 pc size bubbles in the ISM \citep{stine_survey}. 
\par Scintillation arc studies require high S/N to obtain quality data. As a consequence, large surveys of scintillation arcs have typically focused on canonical (i.e., non-recycled) pulsars with high flux densities and low dispersion measures \citep{stine_survey} or observations at low observing frequencies where pulsars are typically brightest \citep{wu_2022}. Some recent large surveys have used newer, more sensitive instruments and observing configurations, in some cases detecting scintillation arcs in over 100 pulsars across a wide range of dispersion measures \citep{2023main_1} and in others performing large scintillation arc surveys using generally lower flux density millisecond pulsars \citep{2023main_2}. Pulsar scintillometry, which uses observations of scintillation arcs over many years, has allowed for high precision estimations on the localization of scattering screens along a given LOS \citep{mult_screen_1508, McKee_2022}. Thanks to long term campaigns to detect low frequency gravitational waves using arrays of millisecond pulsars \citep{Agazie_2023,2023epta,2023ppta,inpta}, measurements of annual arc variations via high precision scintillometry have also recently been accomplished using a few millisecond pulsars \citep{Main, Reardon_2020, 2022mall}.
\par Traditional measurements of scintillation arcs have typically been limited to either one observing band over all epochs (e.g., \cite{trang}), or alternated between observing bands from epoch to epoch (e.g., \cite{stine_ock}). While generally sufficient for most analyses, this band limit results in a bottleneck for examining the evolution of various frequency-dependent effects over shorter timescales, including scintillation arc curvature, structures within individual arcs, and asymmetries in both arc brightness and power as a function of differential time delay. By making use of the subarray capabilities of the upgraded Giant Metrewave Radio Telescope (uGMRT) \citep{gmrt}, we can effectively create an ultra wideband receiver by setting multiple groups of dishes to simultaneously observe at different frequencies. This work is primarily data-focused and aims to highlight the results of some multi-frequency analyses performed on a small survey of six strong canonical pulsars, all known to exhibit scintillation arcs in at least one of the frequency bands, using this approach. In Section \ref{sec:data} we discuss the data taken as part of our survey. Section \ref{sec:analysis} describes the analyses performed and the physical parameters extracted. Section \ref{sec:results} details the results of these analyses. Finally, Section \ref{sec:conclusion} summarizes our results and discusses possible next steps.

\section{Data}
\label{sec:data}
Our data were taken across eight epochs spanning MJD 58987$-$59497 using 22 dishes split into subarrays for simultaneous multi-frequency observations at uGMRT's Band 3 and Band 4, centered at 400 MHz and 650 MHz, respectively, each with 200 MHz of bandwidth. This simultaneous low-frequency accessibility is comparable to instruments like CHIME that can observe continuously between 400$-$800 MHz \citep{chime} and better than instruments such as the Green Bank Telescope, which, while having a wide range of low frequency coverage, can only observe below 1 GHz with at most 240 MHz of bandwidth at frequencies close to 1 GHz and less than 200 MHz of bandwidth in lower frequency ranges \citep{greenbankobservatoryReceiversFrequency}. Observations were also made at Band 5 centered at 1360 MHz, although due to a combination of RFI and low signal-to-noise (S/N) no scintles were detectable in the dynamic spectra. The observing bands were split into 4096 (49 kHz wide) frequency channels and observed with 10 second subintegrations. These data were flux calibrated using observations of either 3C147 or 3C286 taken at the beginning of every observing session, and every pulsar was phase calibrated with a nearby source for five minutes once every 40 minutes of observing time on the pulsar. Two to three pulsars were observed at each epoch for 40 minutes, except for MJD 59497, where three pulsars were observed for 155 minutes each. As a result of the phase calibration, each of those observations comprised three 40 minute scans plus an additional 20 minute scan. A summary of the observations made can be found in Table \ref{pulsar_info}.
\begin{deluxetable}{CCCC}
\tablewidth{0pt}
\tablecolumns{4}
\tablecaption{Pulsars Observed}
\label{pulsar_info}
\tablehead{\multicolumn{2}{C}{\rm Pulsar} & \multicolumn{1}{C}{$N_{\textrm{obs}}$} &    \multicolumn{1}{C}{\rm Frequencies} \\ \multicolumn{1}{C}{\rm J2000 \ Epoch} & \multicolumn{1}{C}{\rm B1950 \ Epoch} & \multicolumn{1}{C}{} & \multicolumn{1}{C}{(\rm MHz)}}
\startdata
\text{J}0630\text{$--$}2834  & \text{B}0628\text{$--$}28 & 1 & 300$-$500, 550$-$750 \\
\text{J}1136{+}1551 & \text{B}1133{+}16 & 4 & 300$-$500, 550$-$750\\
\text{J}1509\text{+}5531 & \text{B}1508\text{+}55 & 4 & 550$-$750\\
\text{J}1645\text{$--$}0317 & \text{B}1642\text{$--$}03 & 1 & 550$-$750\\
\text{J}1932\text{+}1059 & \text{B}1929\text{+}10 & 3 & 300$-$500, 550$-$750\\
\text{J}2048\text{$--$}1616 & \text{B}2045\text{$--$}16 & 1 & 300$-$500, 550$-$750
\enddata
\tablecomments{Summary of pulsar observations.}
\end{deluxetable}

The raw data were reduced using the publically available pipeline 
for uGMRT data reduction, \textsc{PINTA} \citep{smj+21}. Here, first 
radio frequency interference (RFI) was excised from the raw data 
using RFIclean \citep{mvlv21}. Then, the data were folded with the full 
frequency resolution using a pulsar ephemeris obtained from the ATNF pulsar 
catalog \citep{mhth2005} to obtain partially folded sub-banded profiles 
in PSRFITS format \citep{2004PASA}.  All the subsequent analysis used 
these reduced PSRFITS files.

\section{Analysis}
\label{sec:analysis}

All observations were processed to extract their dynamic spectra by calculating the intensity, $S$, of the pulsar’s signal at each observing frequency, $\nu$, and time, $t$, via
\begin{equation}
\label{dynspec}
S(\nu,t)=\frac{P_{\rm{on}}(\nu,t)-P_{\rm{off}}(\nu,t)}{P_{\rm{bandpass}}(\nu,t)},
\end{equation}
where ${P}_{{\rm{bandpass}}}$ is the total power of the observation as a function of observing frequency and time, and ${P}_{{\rm{on}}}$ and ${P}_{{\rm{off}}}$ are the power in all on- and off-pulse components, respectively, as a function of frequency and time. In this process, we reduced the number of pulse profile phase bins from 512 to 64, and defined the on-pulse region as the bins in the summed profile with an intensity $>$5\% of the maximum in a continuous window. Each dynamic spectrum was then broken up into four 50 MHz spectra to allow for more in-depth frequency-dependent analyses and manually zapped of interference by examining dynamic spectra data arrays and removing pixels that were brighter than the brightest scintle maxima. Secondary spectra were then created by taking the squared modulus of the two-dimensional Fourier transform (i.e., the power spectrum) of the corresponding 50 MHz dynamic spectrum and displaying it logarithmically (dB). While this is certainly one of the more commonly used methods for obtaining the secondary spectrum, alternative approaches have recently been explored \citep{alt_sec}. 
\par
\begin{figure}
    \centering
    \includegraphics[scale=0.58]{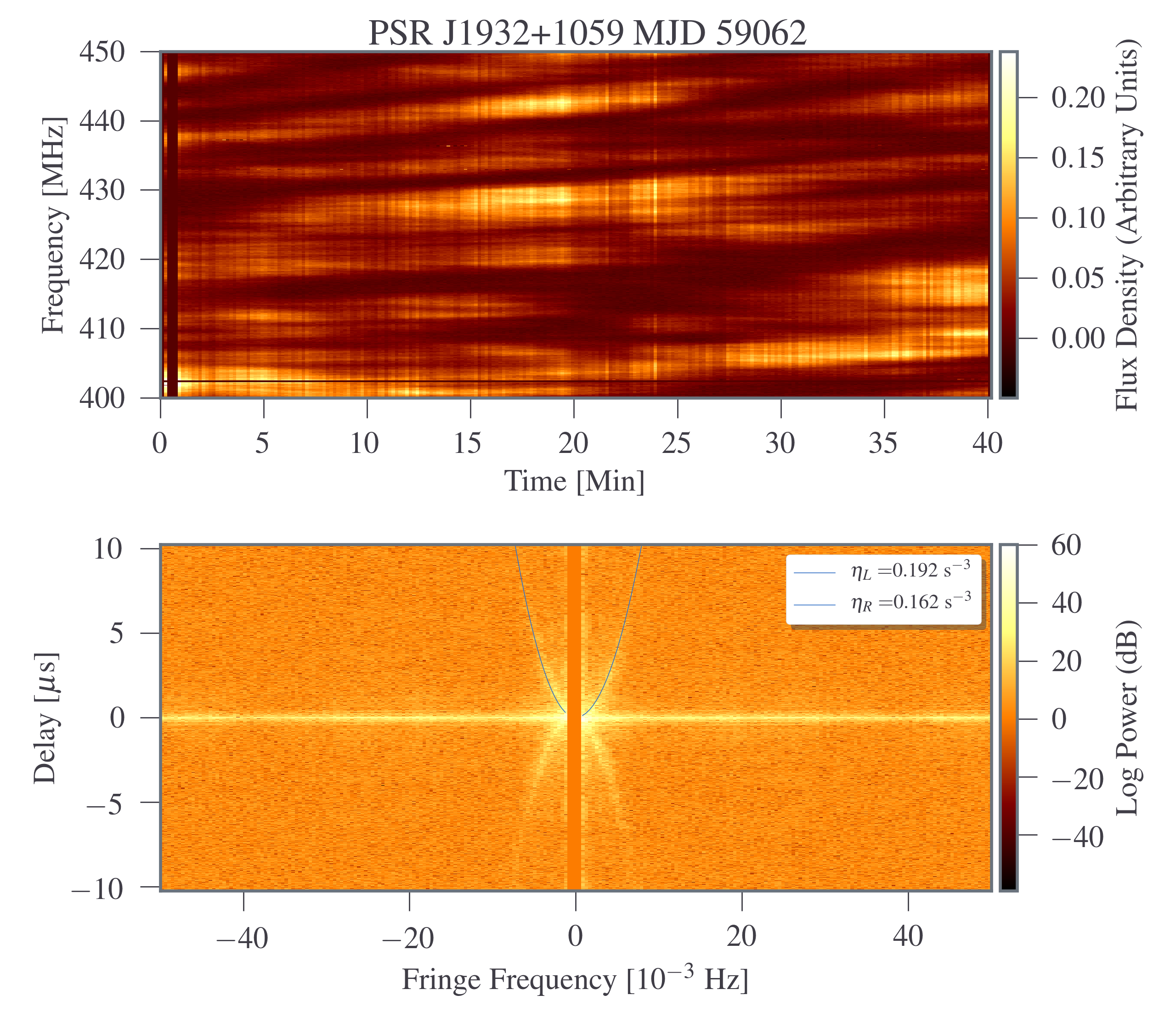}
    \caption{Example dynamic (top) and secondary (bottom) spectra of PSR J1932+1059 centered at 425 MHz on MJD 59062. The top half of the secondary spectrum shows the overlaid arc fits in blue. The complete figure set (104
images) are here attached as .png files in the /anc directory.}
    \label{figset}
\end{figure}

 To determine the arc curvature in the primary (brightest) scintillation arc on both the positive and negative side of each secondary spectrum's fringe frequency axis, we followed the approach described in \cite{stine_survey} in which we divided each secondary spectrum along the center of its fringe frequency axis and further divided the secondary spectrum into horizontal slices up the delay axis until we reached the approximate end of a given arm. Searching only in the region of the spectrum surrounding a given arc, we determined the maximum in each delay slice and fit the resulting trends using $f_{\nu}=\eta f_t^2$ fits, where $f_{\nu}$ is the differential time delay, $\eta$ is the arc curvature, and $f_t$ is the fringe frequency. An example dynamic and secondary spectrum pair with its corresponding primary scintillation arc fits is shown in Figure \ref{figset}. All 104 dynamic and secondary spectra with their corresponding primary scintillation arc fits, displayed as Figure set 1 in the HTML version of the online ApJ version of this paper, are here attached as .png files in the /anc directory. 
\par Normalized secondary spectrum power profiles were acquired using the \textsc{scintools} package \citep{scintools}, which creates these profiles from normalized secondary spectra, which are secondary spectra that have been manipulated such that their scintillation arcs are fully vertical. This package utilizes a Hough transform approach for fitting arcs, in which a range of possible $\eta$ values is explored by calculating the summed power along each corresponding fit, with the resulting curvature of the observation being the one with the greatest summed power.
\par We also determined scintillation parameters by fitting Gaussians and Lorentzians to frequency slices of a given observation's two-dimensional autocorrelation function (ACF) to determine their scintillation bandwidth, $\Delta \nu_{\textrm{d}}$, defined as the half-width at half-maximum (HWHM) of the frequency ACF at lag 0, Gaussians to the time slices of the same ACF to determine their scintillation timescale, $\Delta t_{\textrm{d}}$, defined as the half-width at $e^{-1}$ of the time ACF lag 0, and scintillation drift rate, $\textrm{d}\nu/\textrm{d}t$, defined as the rotation of the 2D Gaussian fit to the 2D ACF in the plane of the frequency and time lags. The decision to measure scintillation bandwidths using both Gaussian and Lorentzian fits serves as a compromise between literature consistency and mathematical rigor; virtually all studies that use an ACF analysis to determine scintillation parameters do so using Gaussian fits. However, given that the ISM's impulse response function, a time domain function, is characterized by a one-sided decaying exponential, it makes more mathematical sense to fit the corresponding frequency domain ACF using a Lorentzian, as the two functions are a Fourier pair. 
\par These scintillation parameter fits were accomplished using code that was heavily based on \textsc{pypulse} \citep{pypulse}, but modified to allow for more user flexibility regarding data ranges over which fits took place, as well as the inclusion of algorithms that allowed for Lorentzian fits.

\section{Results \& Discussion}
\label{sec:results} 
\subsection{Scintillation Arc Curvature Scaling Behavior}
\par As mentioned earlier, \cite{Hill_2003} demonstrated through both theoretical and observational means that the arc curvature $\eta$ should follow a $\nu^{-2}$ dependence, following the same power law as the angular deflection of the pulsar signal. While over 2 GHz of bandwidth was used in those observations (10-12.5 MHz of bandwidth centered at 430 MHz and either 50 or 100 MHz of bandwidth centered at 1175 MHz, 1400 MHz, and 2250 MHz), the frequency coverage was discontinuous and all $\eta$ measurements used in their corresponding fits were from different epochs. Generally the latter point should not be an issue as long as the observations were taken within a period shorter than the pulsar's refractive timescale and the measured effective velocity has changed minimally. Indeed, for the data used in their fits, their measured arc curvatures at a given frequency did not vary significantly on day or week timescales, making them suitable for this type of analysis. However, the ideal situation would be to obtain many measurements at many frequencies during the same observation, preferably at the same time for optimal consistency. With our high resolution and sufficient observing time, we have the ability to make up to eight concurrent arc measurements over 450 MHz of bandwidth at low frequency and can consequentially provide a more definitive examination of the theory. Additionally, since all of our measurements for a given scaling fit are taken on the same day, they all have the same effective velocity.
\par
\begin{figure*}[!ht]
    \centering
    \subfloat[\centering Arc curvature scaling index fit for both left (solid blue with dots) and right (dashed green with triangles) arms for PSR \text{J}1932{+}1059 on MJD 58997  ]{{\includegraphics[width=0.5\textwidth]{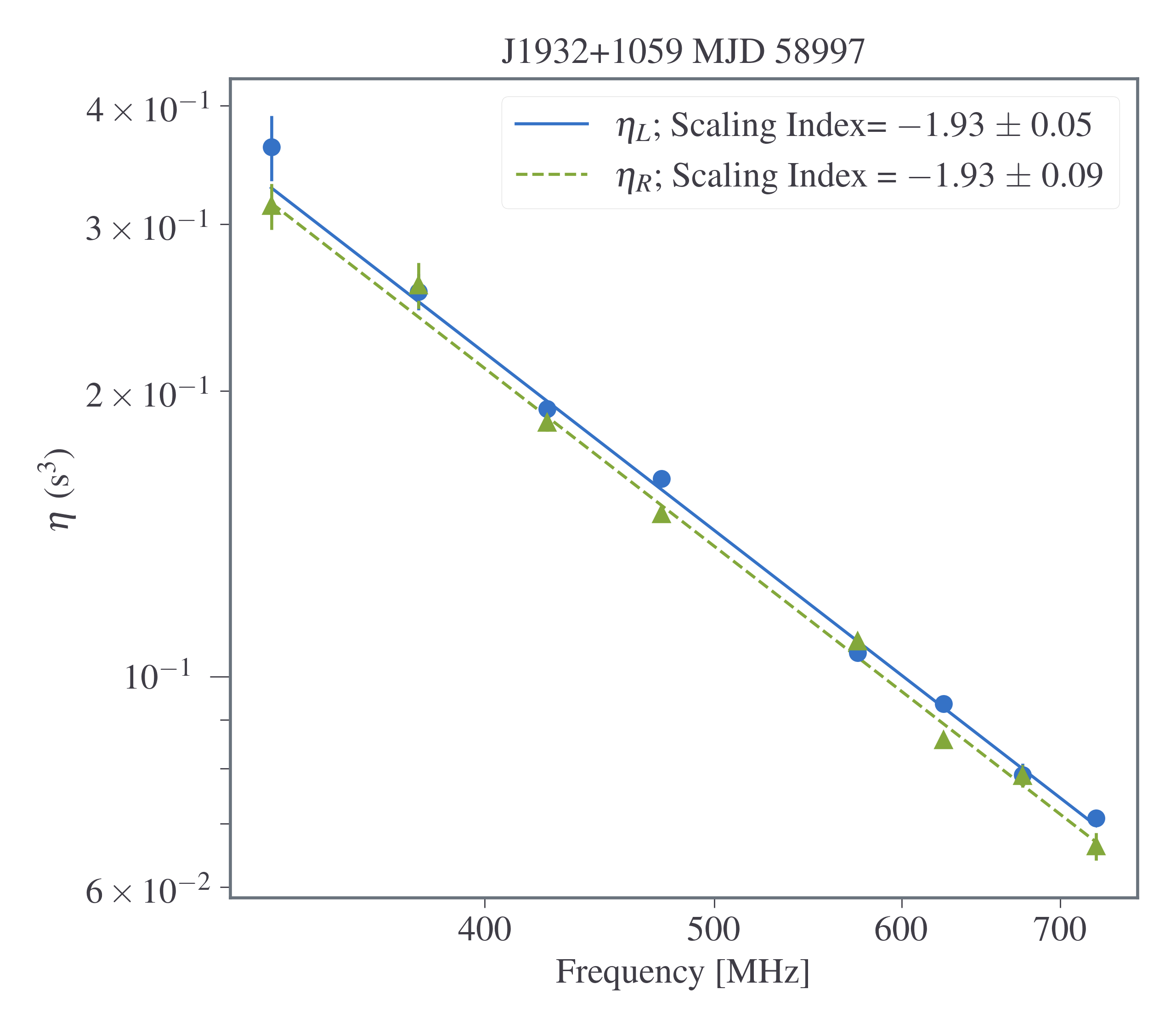} }}%
    \subfloat[\centering Arc curvature scaling index fit both left (solid blue with dots) and right (dashed green with triangles) arms for PSR \text{J}1136{+}1551 on MJD 58997. The inclusion of multiple points at certain frequencies is the result of this epoch containing a 155 minute observation instead of the 40 minutes of the other observations, and so a new $\eta$ was measured after every 40 minutes. ]{{\includegraphics[width=0.5\textwidth]{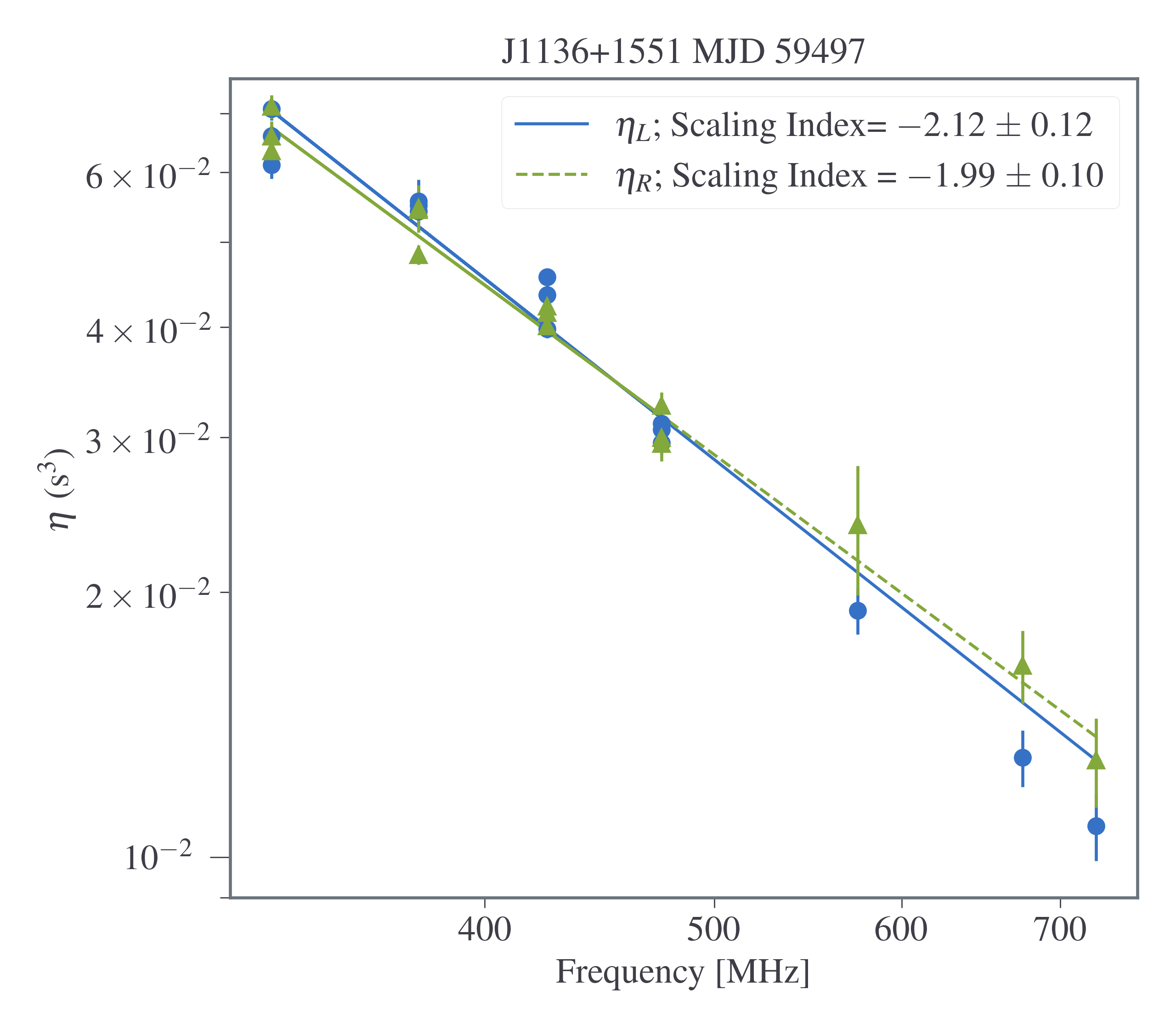} }}%
    \caption{Example fits for the arc curvature scaling index. The complete figure set (14
images) are here attached as .png files in the /anc directory.}%
    \label{ex_scale_fit}%
\end{figure*}

 Following the methodology of \cite{Hill_2003}, for a scaling index $\alpha$, we performed a  weighted linear least-squares fit of the form
\begin{equation}
\label{arc_scale_fit}
    \log_{10}\eta = \alpha\log_{10}\nu+\beta 
\end{equation}
on the curvatures for each pulsar with at least three measurements at a given MJD, weighted by the squared inverses of the arc curvature uncertainties. Example fits can be seen in Figure \ref{ex_scale_fit}, with all measured indices listed in Table \ref{scaling_results}. All 14 scaling fits, displayed as Figure set 2 in the HTML version of the online ApJ version of this paper, are here attached as .png files in the /anc directory. We find that, overall, our scaling indices are consistent with a theoretical index of $-2$, with PSRs \text{J}1136{+}1551 and \text{J}1932{+}1059 being especially consistent. Interestingly, a weighted average of all curvature fits across all pulsars shows that our left arm fits are overall more consistent with an index of $-2$ than our right arms, with a weighted average of $-$1.99$\pm$0.03 across all left arm fits compared with $-$1.69$\pm$0.02 across all right arm fits, indicating that refraction may play a role in how closely arc curvature scales as expected with frequency. However, we strongly emphasize that there is significant variation in curvature indices across pulsars and epochs, and that this is only an average. 

\begin{deluxetable*}{CCCCCCCC}[!ht]
\tablewidth{0pt}
\tablecolumns{6}
\tablecaption{Fitted Pulsar Scintillation Arc Curvature Scaling Indices \label{scaling_results}}
       \tablehead{ \colhead{Pulsar} & \colhead{MJD} & \colhead{Scaling Index Left Arc} & \colhead{Scaling Index Error Left Arc} & \colhead{$N_{\eta}$} & \colhead{Scaling Index Right Arc} & \colhead{Scaling Index Error Right Arc} & \colhead{$N_{\eta}$}}
        \startdata
        \text{J}0630\text{$--$}2834 & 58987 & \textrm{---} & \textrm{---} & \textrm{---} & \text{$--$}2.48 & 0.31 & 8\\ 
        \text{J}1136\text{+}1551 & 58987 & \text{$--$}1.79 & 0.11 & 7 & \text{$--$}1.89 & 0.12 & 7\\ 
        \text{J}1136\text{+}1551 & 58991 & \text{$--$}1.65 & 0.12 & 8 & \text{$--$}1.94 & 0.08 & 8 \\ 
        \text{J}1136\text{+}1551 & 59115 & \text{$--$}1.36 & 0.13 & 8 & \text{$--$}1.52 & 0.15 & 8 \\ 
        \text{J}1136\text{+}1551 & 59497 & \text{$--$}2.12 & 0.12 & 15 &\text{$--$}1.99 & 0.10 & 15 \\
        \text{J}1509\text{+}5531 & 58987 & \text{$--$}1.49 & 0.83 & 3 & \text{$--$}1.34 & 0.55 & 3 \\
        \text{J}1509\text{+}5531 & 59064 & \text{$--$}1.62 & 0.26 & 4 & \text{$--$}1.41 & 0.22 & 4 \\
        \text{J}1509\text{+}5531 & 59115 & \text{$--$}0.93 & 0.21 & 4 &\text{$--$}2.31 & 0.18 & 4 \\ 
        \text{J}1509\text{+}5531 & 59497 & \text{$--$}2.01 & 0.87 & 9 &\text{$--$}1.34 & 0.21 & 9 \\
        \text{J}1645\text{$--$}0317 & 59074 & \text{$--$}2.09 & 0.26 & 4 & \textrm{---} & \textrm{---} & \textrm{---} \\
        \text{J}1932\text{+}1059 & 58997 & \text{$--$}1.93 & 0.05 & 8 &\text{$--$}1.93 & 0.09 & 8 \\
        \text{J}1932\text{+}1059 & 59062 & \text{$--$}2.08 & 0.07 & 8 & \text{$--$}1.75 & 0.07 & 8 \\
        \text{J}1932\text{+}1059 & 59497 & \text{$--$}1.77 & 0.09 & 8 &\text{$--$}1.53 & 0.04 & 8 \\
        \text{J}2048\text{$--$}1616 & 59062 & \text{$--$}2.52 & 0.07 & 6 & \text{$--$}1.68 & 0.05 & 6\\ 
        \enddata
        \tablecomments{Fitted arc curvature scaling indices for both left and right primary arcs. $N_{\eta}$ indicates the number of arc curvature measurements used in each fit. Measurements on MJD 59497 may have $N_{\eta}>8$ due to this epoch being 155 minutes rather than the 40 minutes of the other observations, and so a new $\eta$ was measured after every 40 minutes, although arcs may not have been sufficiently resolved/detected in each 40 minute segment.}
\end{deluxetable*}

\subsection{Scintillation Bandwidth \& Scintillation Timescale Scaling Behavior}
\par Our wide frequency coverage also allowed us to examine the scaling index of scintillation bandwidths and  scintillation timescales. Under the assumption that ISM fluctuations follow behaviors consistent with a Kolmogorov medium and that the subinertial part of the wavenumber spectrum dominates, we should expect that scintillation bandwidths scale with frequency as $\Delta \nu_{\textrm{d}}\propto \nu^{4.4}$  and $\Delta t_{\textrm{d}} \propto \nu^{1.2}$ \citep{Romani,Cordes_1998}. Previous studies examining the scattering indices of various pulsars have done so using a number of methods, including multi-frequency measurements within one week \citep{Krishnakumar_2017}, simultaneous multi-frequency measurements \citep{Bhat_2004,Bansal_2019} (although the former primarily used two, and occasionally three, measurements and the latter was at frequencies less than 100 MHz), splitting up measurements from a single frequency band into multiple subbands \citep{Levin_Scat, Krishnakumar_2019, turner_scat}, and using measurements from many epochs taken at two observing bands non-simultaneously \citep{turner_scat}. Since more measurements and more frequency coverage in a single epoch is ideal, the method used in \cite{Bhat_2004} and \cite{Bansal_2019} is the most preferred of the three. The method in this paper utilizes a combination of this approach and the subband approach to maximize the number of delay measurements per epoch, which can be done thanks to our high frequency resolution and sensitivity in both observing bands. Not as many studies exist that examine scintillation timescale frequency scaling, in part due to the longer observing times required to measure these timescales in most pulsars.  Most studies that do explore this behavior use measurements taken from disparate observations \citep{timescale_long}, with those that have made measurements simultaneously at multiple frequencies being limited to ranges at much lower frequencies \citep{Bhat_2018}.
\par
\begin{figure*}[!ht]
    \centering
    \subfloat[\centering Scintillation bandwidth scaling index fit for PSR \text{J}1136{+}1551 on MJD 58991]{{\includegraphics[width=0.5\textwidth]{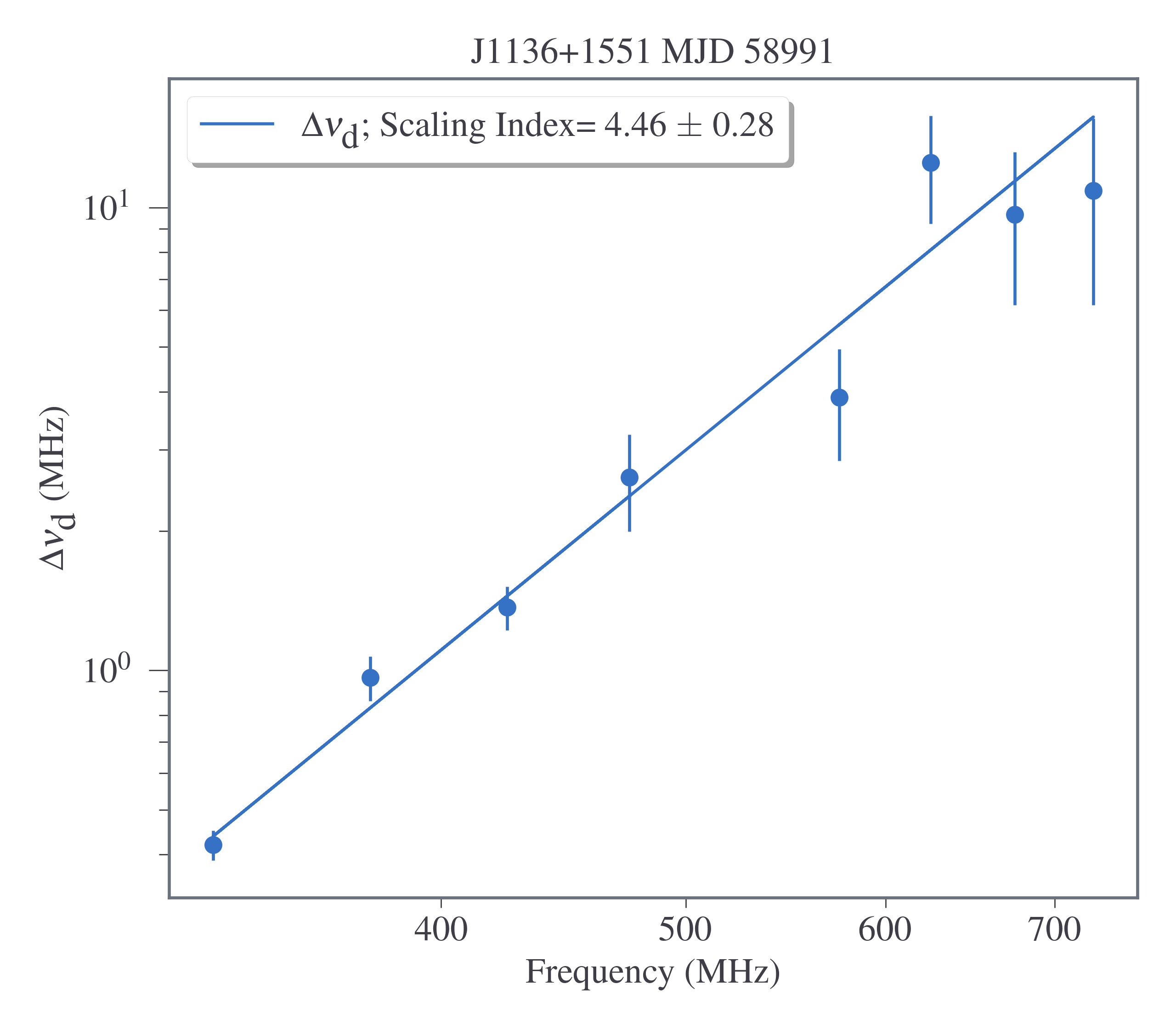} }}%
    \subfloat[\centering Scintillation timescale scaling index fit for PSR \text{J}1932{+}1059 on MJD 59062]{{\includegraphics[width=0.5\textwidth]{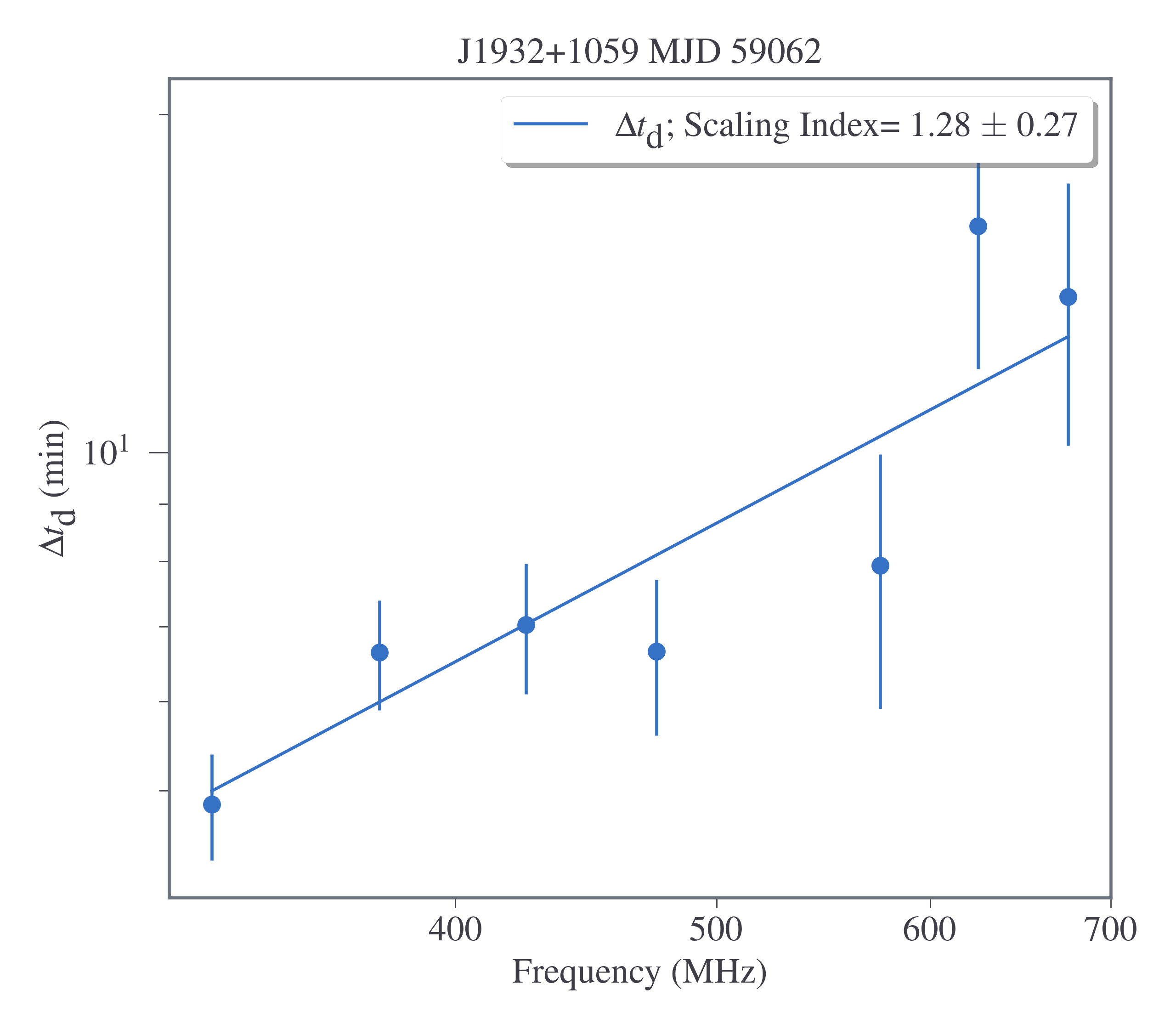} }}%
    \caption{Example scaling index fits. The complete figure set (39
images) are here attached as .png files in the /anc directory.}%
    \label{scatt_scale_ex}%
\end{figure*}

 Similar to Equation \ref{arc_scale_fit} that is used to determine the arc curvature scaling index, our scintillation bandwidth and scintillation timescale scaling indices $\xi$ at each epoch were determined by performing a weighted linear least-squares fit of the form
\begin{equation}
\label{scattering_index_fit}
\log_{10}\Delta \nu_{\textrm{d}}=\xi\log_{10}\nu+b
\end{equation}
and 
\begin{equation}
\label{timescale_index_fit}
\log_{10}\Delta t_{\textrm{d}}=\xi\log_{10}\nu+b
\end{equation}
for pulsars with at least four measurements at a given MJD. Example fits can be seen in Figure \ref{scatt_scale_ex}, with all measured indices listed in Table \ref{scattering_scaling_results}. All 39 scaling fits, displayed as Figure set 3 in the HTML version of the online ApJ version of this paper, are here attached as .png files in the /anc directory. We find that the majority of our scintillation bandwidth indices fit with Gaussians are consistent with scaling shallower than characteristic of a Kolmogorov medium, while half of our scintillation bandwidth indices fit with Lorentzians are consistent with scaling for a Kolmogorov medium. We also find half of our scintillation timescale indices to be consistent with a Kolmogorov medium and half to be consistent with a shallower spectrum. This behavior agrees well with general trends seen in previous studies, as both \cite{Bhat_2004} and \cite{Bansal_2019} found indices either consistent with a Kolmogorov medium or a shallower index than a Kolmogorov medium using delays measured by fitting pulse broadening functions, while \cite{Levin_Scat} and \cite{turner_scat} only found indices that were shallower than a Kolmogorov medium using Gaussian fits to frequency ACFs. \cite{Krishnakumar_2017} measured delays in 47 pulsars over a range of frequencies by fitting pulse broadening functions to pulse profiles and found almost 65\% of those pulsars to exhibit scaling indices shallower than $-4.4$. Additionally, while Gaussians have been used to fit frequency ACFs in many scintillation studies, which is why we include them in our analyses, it is more mathematically appropriate to use Lorenztians given the Fourier relationship between the Lorentzian distribution and the one-sided exponential function that is assumed to characterize the impulse response function of the ISM. As such, our scintillation bandwidth indices obtained using Lorentzian-derived scintillation bandwidths likely provide a more accurate reflection of the turbulence of the ISM than our Gaussian measurements.
\par Many explanations have been given for why shallower-than-Kolmogorov medium behavior has been observed so frequently. Physical arguments have called into question the validity of the simple infinite, thin screen model, demonstrating that shallower scaling indices are more consistent with finite, thin screens \citep{Rickett_2009}. This is expected to be much more common among low DM pulsars \citep{Cordes_2001}, which agrees with our results, as all of the pulsars have dispersion measures below 40 pc cm$^{-3}$. Shallower indices have also been attributed to the existence of multiple finite screens along the LOS \citep{Lew_1}. This hypothesis agrees well with our measured indices for PSR B1133+16, as its indices are consistently shallower than that of a Kolmogorov medium and it is also known to have at least six distinct scattering screens \citep{McKee_2022}. 
\par Quality-of-data arguments have also been proposed. \cite{turner_scat} suggested their shallower indices may be at least partially attributable to an imbalance of lower frequency data to higher frequency data for their multiple epoch approach as well as a lack of sufficient frequency resolution in their lower frequency band in some epochs. However, neither of these issues should affect our results, as our observations have a consistently even balance of low and high frequency measurements at all epochs and all of our measurements are well-resolved in frequency. 

\begin{deluxetable*}{CCCCCCCCCC}[!ht]
\tablewidth{0pt}
\tablecolumns{10}
\tablecaption{Fitted Pulsar Scintillation Bandwidth \& Scintillation Timescale Scaling Indices     \label{scattering_scaling_results}}
       \tablehead{ \colhead{Pulsar} & \colhead{MJD} & \colhead{$\Delta \nu_{\textrm{d}}$ Gaussian} & \colhead{$\Delta \nu_{\textrm{d}}$ Gaussian Error} & \colhead{$\Delta \nu_{\textrm{d}}$ Lorentzian} & \colhead{$\Delta \nu_{\textrm{d}}$ Lorentzian Error} & \colhead{$N_{\Delta \nu_{\textrm{d}}}$} & \colhead{$\Delta t_{\textrm{d}}$ Index} & \colhead{$\Delta t_{\textrm{d}}$ Index Error} & \colhead{$N_{\Delta t_{\textrm{d}}}$}}
       \startdata
        \text{J}0630\text{$--$}2834 & 58987 & 4.85 & 0.66 & 4.85 & 0.65 & 8 & 1.02 & 0.25 & 8\\ 
        \text{J}1136{+}1551 & 58987 & 2.63 & 0.78 & 3.47 & 1.05 & 7 & 1.51 & 0.21 & 7\\ 
        \text{J}1136{+}1551 & 58991 & 3.74 & 0.25 & 4.46 & 0.28 & 8 & 0.70 & 0.12 & 8\\
        \text{J}1136{+}1551 & 59115 & 2.72 & 0.35 & 2.87 & 0.37 & 8 & 0.56 & 0.27 & 8\\
        \text{J}1136{+}1551 & 59497 & 3.51 & 0.66 & 4.07 & 0.58 & 14 & 0.71 & 0.16 & 14\\
        \text{J}1509{+}5531 & 59064 & 0.56 & 1.39 & 0.42 & 1.26 & 4 & 0.40 & 0.38 & 4\\
        \text{J}1509{+}5531 & 59115 & 1.87 & 0.81 & 1.09 & 1.53 & 4 & 1.22 & 0.69 & 4\\
        \text{J}1509{+}5531 & 59497 & 0.73 & 0.81 & 0.38 & 1.03 & 9 & \text{$--$}0.58 & 0.48 & 9\\
        \text{J}1645\text{$--$}0317 & 59074 & 3.70 & 0.80 & 3.55 & 0.75 & 8 & 0.55 & 0.06 & 8 \\ 
        \text{J}1932{+}1059 & 58997 & 1.89 & 0.46 & 2.47 & 0.48 & 8 & \text{$--$}0.28 & 0.49 & 8 \\ 
        \text{J}1932{+}1059 & 59062 & 2.02 & 0.42 & 2.53 & 0.46 & 8 & 1.28 & 0.27 & 7\\ 
        \text{J}1932{+}1059 & 59497 & 3.77 & 0.34 & 4.37 & 0.54 & 8 & 0.89 & 0.57 & 8\\
        \text{J}2048\text{$--$}1616 & 59062 & 3.75 & 0.33 & 4.21 & 0.44 & 8 & 0.37 & 0.08 & 6\\
        \enddata
        \tablecomments{Fitted scintillation bandwidth and scintillation timescale scaling indices, with a minimum of four measurements ($N$) required in a given epoch to obtain a scaling index. Errors are uncertainties from parameter fits. The majority of our scintillation bandwidth indices fit with Gaussians are consistent with scaling shallower than characteristic of Kolmogorov medium, while the majority of our scintillation bandwidth indices fit with Lorentzians are consistent with scaling for a Kolmogorov medium. Measurements on MJD 59497 may have $N>8$ due to this epoch being 155 minutes rather than the 40 minutes of the other observations, and so a new $\eta$ was measured after every 40 minutes, although arcs may not have been sufficiently resolved/detected in each 40 minute segment.}
\end{deluxetable*}

\subsection{Relation Between Arc Curvature and Scintillation Parameters}
Under the assumption that scattering resulting in a given scintillation arc is both one-dimensional and occurring at a thin screen, it becomes possible to relate arc curvature to the scintillation bandwidth and timescale \citep{cordes_2006_refraction, stine_survey}. This relation can be described by the so-called pseudo-curvature, $\eta_{\rm ISS}$, given by 
\begin{equation}
    \eta_{\rm ISS} = \frac{2\pi\Delta t_{\rm d}^2}{\Delta \nu_{\rm d}}.
\end{equation}

\begin{figure}[!ht]
    \centering
   \includegraphics[width=0.5\textwidth]{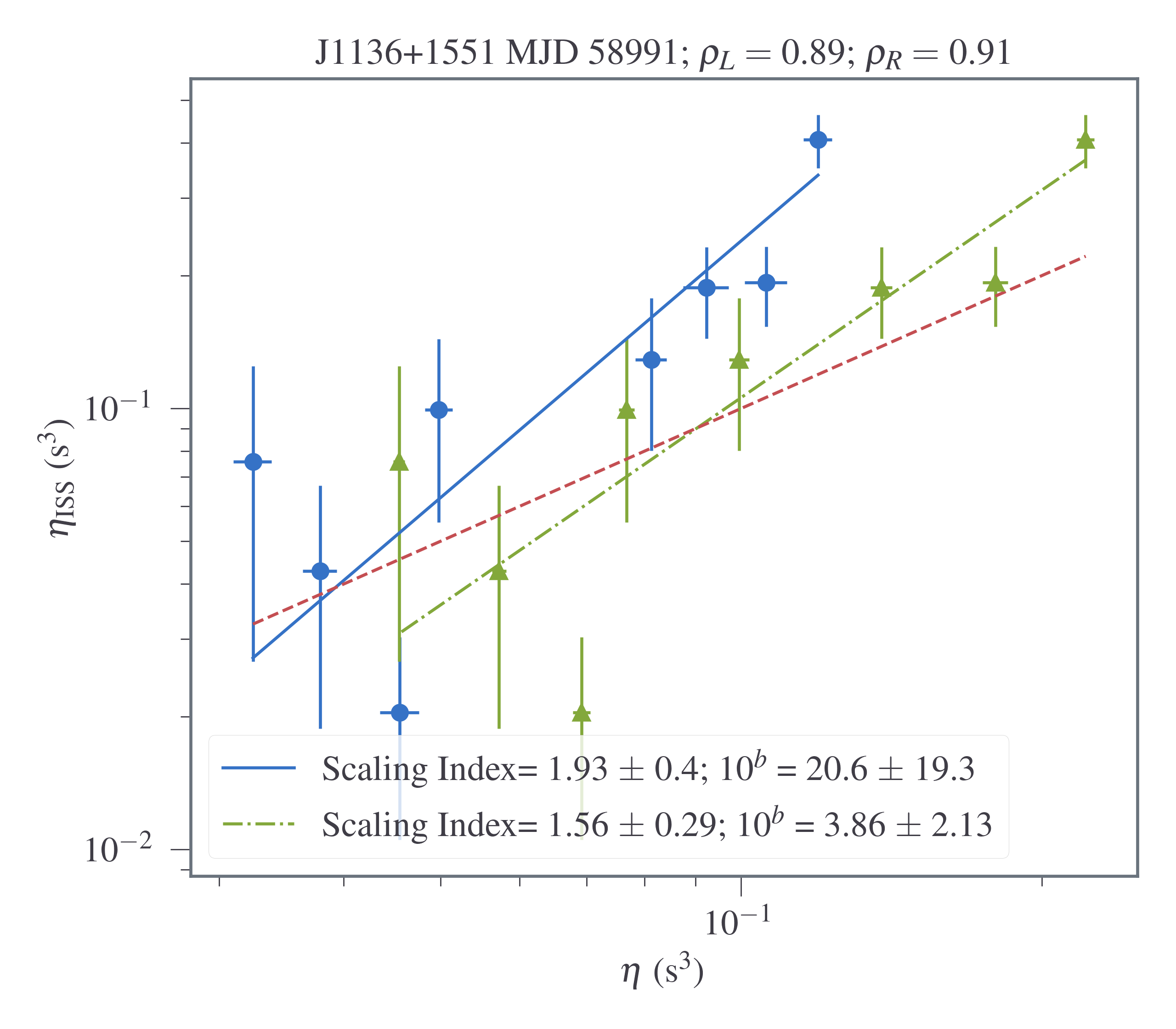} %
    \caption{Example comparison between $\eta_{\rm ISS}$ and $\eta$ for PSR J1136+1551 on MJD 58991 for both the left (solid blue with dots) and right (dotted-dashed green with triangles) arms of the scintillation arc visible in the observation, with corresponding power law fits, and correlation coefficients $\rho_{\rm L}$ and $\rho_{\rm R}$, respectively. The dashed red line indicates a perfect one-to-one correspondence.}%
    \label{eta_iss_ex}%
\end{figure}

While the relation between $\eta_{\rm ISS}$ and $\eta$ has been strongly demonstrated in large surveys across many pulsars \citep{stine_survey}, our wide frequency coverage and high resolution uniquely allow for the exploration of this relation within individual epochs. To accomplish this, in each epoch, for a given pulsar that had at least four measurements of arc curvature in a given arm and four measurements of both scintillation bandwidth and timescale, we determined the weighted linear correlation coefficient in log space between $\eta_{\rm ISS}$ and $\eta$. We also determined the power law relation between $\eta_{\rm ISS}$ and $\eta$ using relations of the same form as Equations \ref{scattering_index_fit} and \ref{timescale_index_fit} to determine the degree of one-to-one correspondence, with a scaling index of one and $10^b$ of one indicating perfect equivalence. The results of these analyses can be seen in Table \ref{eta_iss_results}, while an example can be seen in Figure \ref{eta_iss_ex}.

\begin{figure}[!ht]
    \centering
   \includegraphics[width=0.5\textwidth]{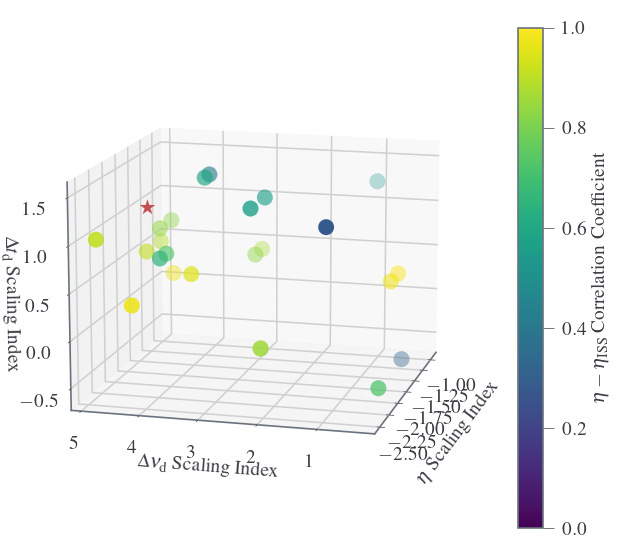} %
    \caption{A comparison of the measured $\eta$, $\Delta \nu_{\rm d}$, and $\Delta t_{\rm d}$ power laws with the correlation coefficients of the $\eta-\eta_{\rm ISS}$ relation (dots), with the red star indicating a power law in $\eta$ of $-2$, a power law in $\Delta \nu_{\rm d}$ of 4.4, and a power law in $\Delta t_{\rm d}$ of 1.2. There is minor evidence that, the closer these three quantities get to their expected scaling indices, the closer the correlation coefficient gets to one.}%
    \label{3d_compare}%
\end{figure}

Overall, we found strong correlations between these two estimates of curvature, with the majority of correlation coefficients being above 0.8, and the majority of fits closely following a one-to-one correspondence, with a weighted average index of $1.16 \pm 0.05$ and weighted average $10^b$ of $0.13 \pm 0.01$, indicating our assumption of one-dimensional, thin-screen scattering is justified across these LOS's. A comparison of the measured $\eta$, $\Delta \nu_{\rm d}$, and $\Delta t_{\rm d}$ power laws in 3D space using the corresponding correlation coefficients as a gauge shows some evidence that these correlation coefficients generally increase the closer all three power laws converge to their expected scaling indices given our various physical assumptions, as shown in Figure \ref{3d_compare}.

\begin{deluxetable*}{CCCCCCCCCC}[!ht]
\tablewidth{0pt}
\tablecolumns{6}
\tablecaption{Relation Between $\eta$ and $\eta_{\rm ISS}$ \label{eta_iss_results}}
       \tablehead{ \colhead{Pulsar} & \colhead{MJD}  & \colhead{$\rho_{\rm L}$} & \colhead{Left Index} & \colhead{Left $10^b$} & \colhead{$N_{\eta, \rm{L}}$} & \colhead{$\rho_{\rm R}$} & \colhead{Right Index} & \colhead{Right $10^b$} & \colhead{$N_{\eta, \rm{R}}$}}
        \startdata
        \text{J}0630\text{$--$}2834 & 58987 & \textrm{---} & \textrm{---} & \textrm{---} & \textrm{---} & 0.91 & 1.35 $\pm$ 0.26 & 0.72 $\pm$ 0.12 & 8 \\
        \text{J}1136{+}1551 & 58987 & 0.42  & 0.56 $\pm$ 0.60 & 0.38 $\pm$ 0.50 & 7 & 0.60 & 0.61 $\pm$ 0.41 & 0.36 $\pm$ 0.28 & 7 \\
        \text{J}1136{+}1551 & 58991 & 0.89 & 1.93 $\pm$ 0.40 & 20.6 $\pm$ 19.3 & 8 & 0.91 & 1.56 $\pm$ 0.29 & 3.89 $\pm$ 2.13 & 8 \\
        \text{J}1136{+}1551 & 59115 & 0.86 & 1.24 $\pm$ 0.30 & 2.70 $\pm$ 2.39 & 8 & 0.83 & 1.15 $\pm$ 0.31 & 1.64 $\pm$ 1.38 & 8 \\
        \text{J}1136{+}1551 & 59497 & 0.66 & 1.12 $\pm$ 0.37 & 2.77 $\pm$ 3.04 & 14 & 0.73 & 1.30 $\pm$ 0.36 & 4.80 $\pm$ 5.09 & 14 \\
        \text{J}1509{+}5531 & 59064 & 0.98 & 1.03 $\pm$ 0.14 & 0.13 $\pm$ 0.01 & 4 & 0.99 & 1.21 $\pm$ 0.15 & 0.14 $\pm$ 0.01 & 4 \\
        \text{J}1509{+}5531 & 59115 & 0.52 & 0.52 $\pm$ 0.61 & 0.10 $\pm$ 0.09 & 4 & 0.28 & 0.12 $\pm$ 0.29 & 0.06 $\pm$ 0.02 & 4 \\
        \text{J}1509{+}5531 & 59497 & 0.72 & 0.70 $\pm$ 0.25 & 1.23 $\pm$ 0.51 & 9 & 0.31 & 0.67 $\pm$ 0.78 & 0.96 $\pm$ 0.97 & 9 \\
        \text{J}1645\text{$--$}0317 & 59074 & 0.95 & 1.14 $\pm$ 0.26 & 0.95 $\pm$ 0.39 & 4 & \textrm{---} & \textrm{---} & \textrm{---} & \textrm{---} \\
        \text{J}1932{+}1059 & 58997 & 0.84 & 1.48 $\pm$ 0.39 & 6.44 $\pm$ 4.62 & 8 & 0.87 & 1.56 $\pm$ 0.37 & 8.09 $\pm$ 5.65 & 8 \\
        \text{J}1932{+}1059 & 59062 & 0.57 & 0.90 $\pm$ 0.76 & 1.97 $\pm$ 2.25 & 7 & 0.55 & 0.91 $\pm$ 0.80 & 2.39 $\pm$ 3.23 & 7 \\
        \text{J}1932{+}1059 & 59497 & 0.84 & 2.17 $\pm$ 0.59 & 52.0 $\pm$ 53.9 & 8 & 0.87 & 2.45 $\pm$ 0.69 & 128 $\pm$ 168 & 8 \\
        \text{J}2048\text{$--$}1616 & 59062 & 0.97 & 1.36 $\pm$ 0.18 & 3.26 $\pm$ 1.24 & 6 & 0.98 & 2.01 $\pm$ 0.23 & 39.3 $\pm$ 23.3 & 6
        \enddata
        \tablecomments{Linear correlation coefficients, $\rho$, and power law fits in log space between measured arc curvature $\eta$ and pseudo-curvature $\eta_{\rm ISS}$ for both left and right arms of scintillation arcs. $N_{\eta}$ indicates the number of arc curvature measurements used in each fit. Measurements on MJD 59497 may have $N_{\eta}>8$ due to this epoch being 155 minutes rather than the 40 minutes of the other observations, and so a new $\eta$ was measured after every 40 minutes, although arcs may not have been sufficiently resolved/detected in each 40 minute segment.}
\end{deluxetable*}

\subsection{155 Minute Observation}
\par The inclusion of a 155 minute observation in our survey on MJD 59497 allowed for an analysis of short-term arc curvature variation in some pulsars, as observations had to be paused every 40 minutes for a five minute phase calibration, resulting in multiple 40 minute scans. For pulsars with at least two arc curvature measurements in a given scintillation arc at a given observing frequency, we examined how a given arc curvature measurement at a given observing frequency, $\eta_{\nu,i}$ varied with respect to the weighted average curvature for that arm and frequency over the entire epoch, $\overline{\eta_{\nu}}$. The percent difference $\chi$ between these two quantities is then given by 
    \begin{equation}
        \chi =  100\frac{|\overline{\eta_{\nu}}-\eta_{\nu,i}|}{\frac{\overline{\eta_{\nu}}+\eta_{\nu,i}}{2}}.
    \end{equation}
\par For PSR J1136{+}1551, all observing frequencies centered at or below 475 MHz had three arc curvature measurements (the higher frequency observations only had one arc curvature measurement) in each primary arm (the brightest arm, overwhelmingly often the arm with the highest curvature) at each frequency, with the accumulation of all percent differences yielding a bimodal distribution with peaks around percent differences of 2\% (16 arc curvature measurements) and 7\% (eight arc curvature measurements). The mean was 3.7\% and the median was 2.8\%, while largest percent difference away from a weighted mean was 7.9$\pm$0.2\% and the smallest was 0.14$\pm$1.91\%, although the majority of all percent differences were below 3\%. All of this strongly indicates the ISM underwent very little change along the LOS to this pulsar over the course of a given observation. This result is supported by this pulsar's incredibly low dispersion measure, meaning it does not sample a sizeable portion of the ISM along its LOS relative to many pulsars that are observed \citep{atnf_1,atnf_3,atnf_4}. 
\par For PSR J1509{+}5531, all observing frequencies centered at or above 575 MHz had at least two arc curvature measurements in each arm at each frequency (no arcs were sufficiently resolved in the lower frequency band), with the accumulation of all percent differences resulting in a one-sided distribution peaked around 6\% (18 arc curvature measurements). The mean was 10.1\% and the median was 6.6\%, while the smallest percent difference away from a weighted mean was 1.2$\pm$2.1\%, and the largest was 36.4$\pm$0.1\%, although the next largest after that was only 21.7$\pm$0.1\%, meaning this maximum was an extreme outlier. The majority of all percent differences were below 7\%. As with the previous pulsar, this also strongly indicates the ISM underwent very little change along the LOS to this pulsar over the course of a given observation, a result again supported by this pulsar's fairly low dispersion measure \citep{atnf_2,atnf_3}. The fact that this pulsar shows higher variation compared to PSR J1136{+}1551 is likely due to PSR J1509{+}5531 having a dispersion measure four times higher and a transverse velocity 45\% larger \citep{atnf_1,atnf_2,atnf_3, atnf_4,atnf_5}, so a significantly larger fraction of the ISM was sampled during its observation, increasing the likelihood of larger scintillation-based variations.

Only one arc curvature measurement in each frequency was obtainable at this MJD for PSR J1932+1059, and so the above analysis was not possible. 

\bigbreak
\par The next few subsections highlight the features of a few pulsars in the survey.

\subsection{J0630-2834}

\begin{figure}
    \centering
    \includegraphics[scale = 0.58]{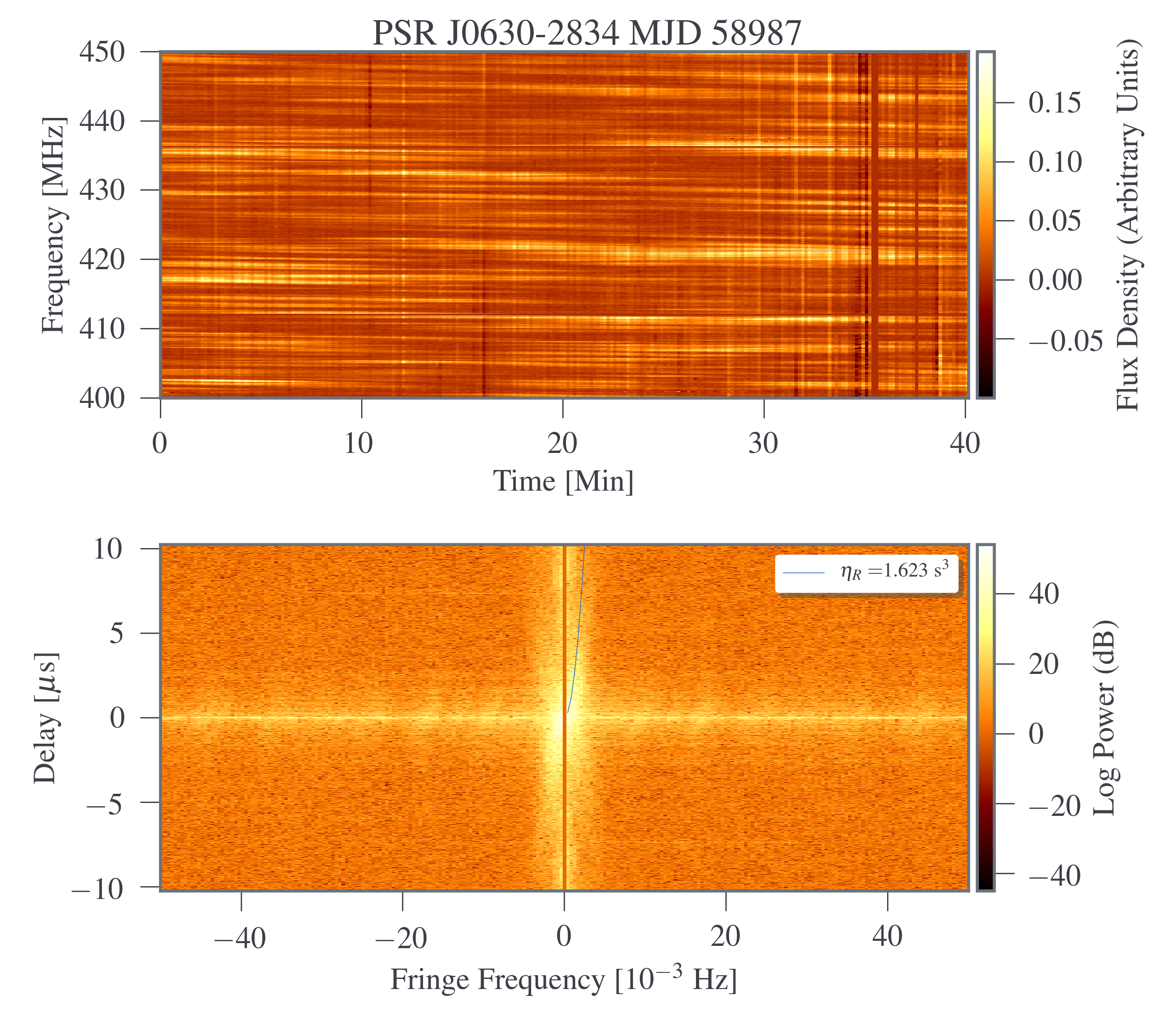}
    \caption{An example dynamic (top) and secondary (bottom) spectrum from PSR J0630-2834 on MJD 58987 centered at 425 MHz. There is a clear asymmetry in the secondary spectrum, with the right arm being the dominant feature. This is likely the result of refraction along the line of sight. The blue line represents the arc curvature fit.\iffalse Scaled uncertainties of the arc curvature can be found in Table \ref{gen_results}.\fi}
    \label{0630_ex}
\end{figure}

In the one epoch for which we were able to resolve a scintillation arc, only the right arm was resolvable across all frequencies, with its relative brightness with respect to the left side of the fringe frequency axis consistently decreasing as frequency increased. An example of this asymmetry can be seen in Figure \ref{0630_ex}. This strong asymmetry is known to be the result of refraction leading to scintillation drifting in the dynamic spectra \citep{cordes_2006_refraction}. Interestingly, although one would expect an increase in scintillation drift to coincide with an increase in the asymmetry, the magnitude of our measured scintillation drift rates seem to mildly favor an increase with frequency whereas the asymmetry appears to decrease with frequency.  

\subsection{J1136+1551}
\begin{figure*}[!ht]
    \centering
    \subfloat[\centering Scintillation arcs without overlaid fits]{{\includegraphics[width=0.5\textwidth]{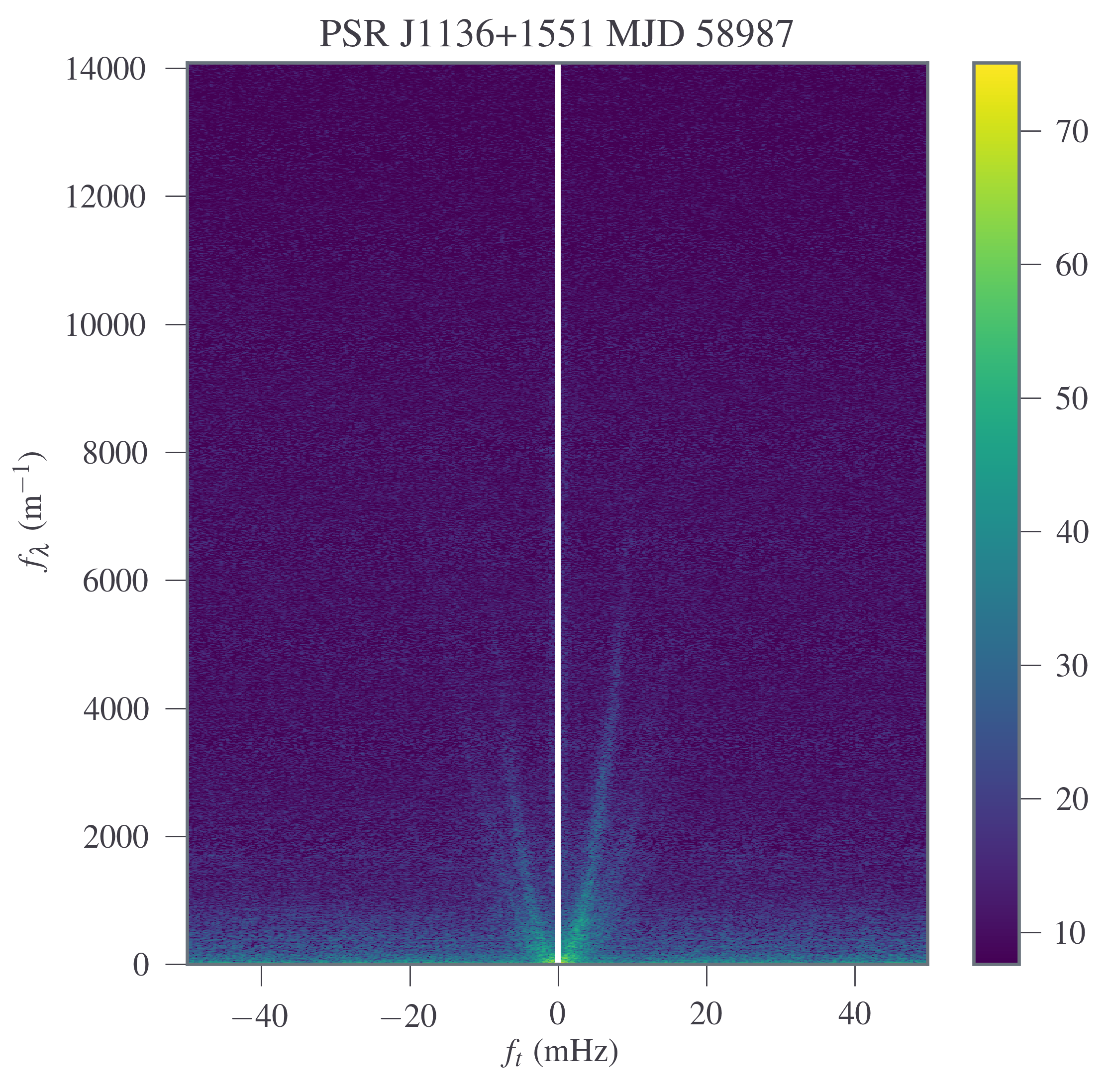} }}%
    \subfloat[\centering Scintillation arcs with overlaid fits]{{\includegraphics[width=0.5\textwidth]{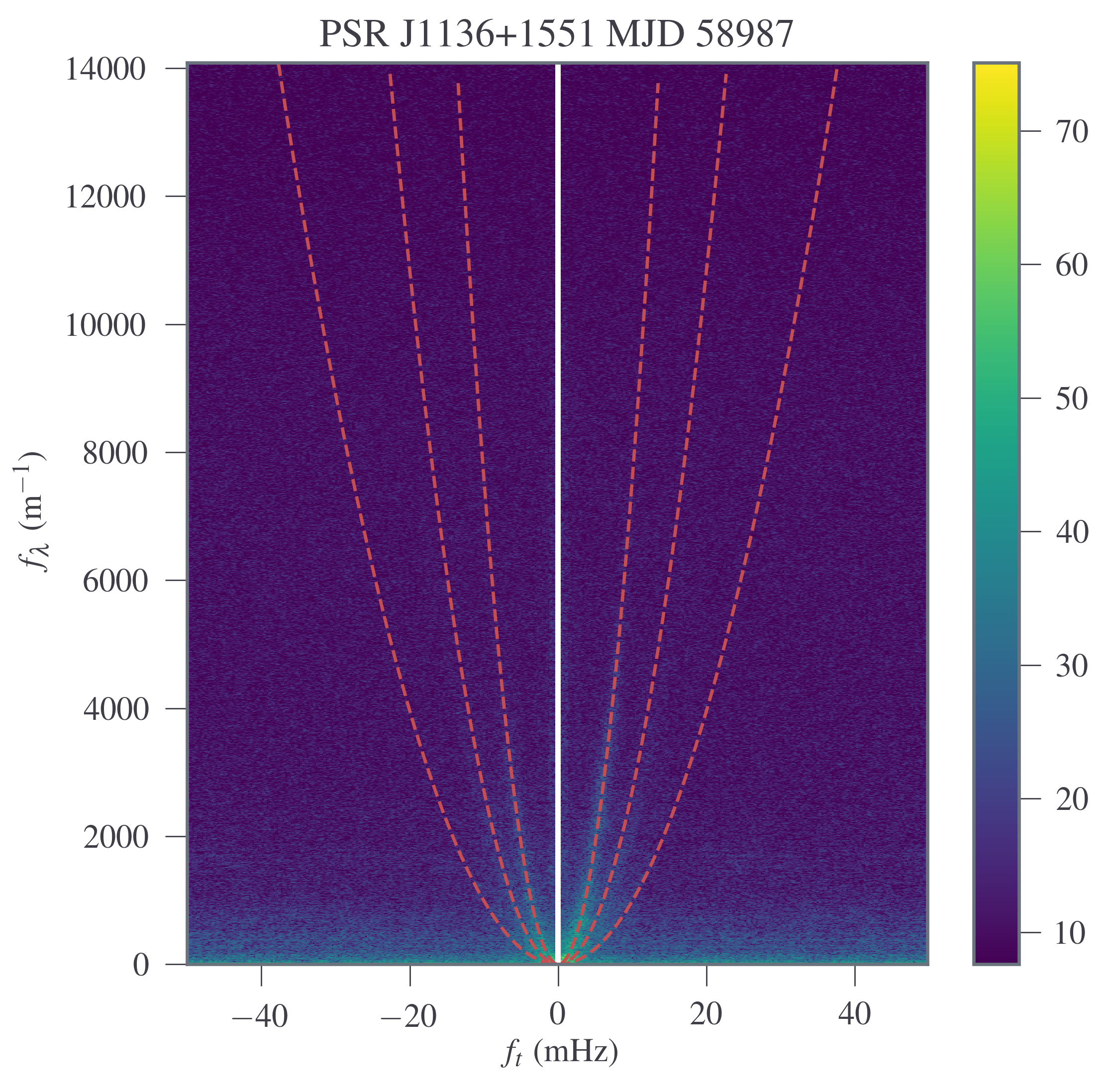} }}%
    \caption{Secondary spectrum of PSR J1136+1551 at 650 MHz on MJD 58987 showing the detection of three distinct scintillation arcs. The dynamic spectrum used to make these figures was resampled to be uniform in wavelength rather frequency in order to better resolve the additional distinct arcs, which were not quite as resolved using the latter sampling approach.}%
    \label{mult_arc}
\end{figure*}

This pulsar is well known for having the uncommon feature of multiple scintillation arcs, implying multiple scattering screens along its LOS \citep{Hill_2003,stine_ock}. In the literature, six distinct sets of arcs have been found over a $\sim$34 year span of observations \citep{McKee_2022}. In three of the four epochs in which we observed this pulsar, we detected multiple arcs, an example of which can be seen in Figure \ref{mult_arc}. 
After scaling our measurements to 1400 MHz and using the convention from \cite{McKee_2022}, we can conclude that we detected arcs B, C, and E on MJDs 58987 and 58991 and arcs C and D on MJD 59115, with arc C being the only detectable arc on MJD 59497. All multiple-arc detections were made only in the observations using uGMRT's band 4, which was centered at 650 MHz. The fact that the two epochs closest to each other in our survey (MJDs 58987 and 58991) both detected the same sets of arcs may hint at a timescale over which certain screens have a larger influence over the pulsar signal propagation. 
\cite{McKee_2022} demonstrates how for this pulsar anywhere from one to six arcs may be visible in a given observation, with the number of visible arcs possibly changing on approximately week timescales, assuming the ISM along the LOS to this pulsar is consistent modulo one year. This, along with our results showing observations within a four day window to be dominated by the same screen, suggests a fairly short but currently unknown window within which the influence of a given screen will fluctuate, likely on the order of one week.
\par The existence of multiple screens along the LOS to this pulsar may also help to partially explain discrepancies between our scintillation bandwidth power-law fits and those of \cite{wu_2022}, where two of our four Gaussian ACF fit-derived scaling indices are shallower than the index acquired from their observation. While we observed anywhere from one to three arcs in our observations, indicating anywhere from one to three scattering screens having a dominant influence over the pulsar emission's propagation, their observation of this pulsar only had one visible arc and therefore one dominant scattering screen. As discussed earlier, \cite{Lew_1} mentions that deviations from a single thin screen scattering geometry (e.g., multiple thin screens) can result in scaling indices shallower than $-$4. The observations in \cite{wu_2022} are also taken at significantly lower observing frequencies (148$-$152 MHz), where it is much more likely for only single arcs to be observed. This is a consequence of the ISM looking more like a continuous medium when observed at lower frequencies, whereas at higher frequencies we are more likely to observe single scatterings off of individual screens.
\par An examination of the power in each of the arms show notable levels of asymmetry along the delay axis, and consequently a notable amount of refraction, in all detectable arms and across all frequencies in the first two epochs, with the right arm having more power and extending further out on the delay axis. This asymmetry clearly decreases over the course of our observations across all frequencies until our final observation, where the arcs have approximately even levels of power or the left arc starts to dominate in the asymmetry. This trend is generally supported by the measured scintillation drift rates as well, especially for data taken at band 4 (650 MHz), i.e., the same band where the multiple arcs were visible, as measured drifts are generally negative during the first three epochs and then considerably positive during the final epoch.
\par Perhaps the most interesting finding from our observations of this pulsar is the discovery of a strong correlation between the measured arc curvatures and the arc asymmetry index, $A(f_{\nu})$, which is a metric that describes the relative power between the left and right arms and is found by comparing the average power along each arm via
\begin{equation}
A(f_{\nu})=\frac{\overline{P_{R}(f_{\nu})} - \overline{P_{L}(f_{\nu})}}{\overline{P_{R}(f_{\nu})} + \overline{P_{L}(f_{\nu})}},   
\end{equation} 
with a larger index magnitude indicating greater asymmetry. We believe this correlation has never before been reported and wish to use it as a cautionary tale for those attempting similar analyses in the future, as we suspect this correlation to be a consequence of a bias in the Hough transform approach used to acquire these asymmetry measurements. To test our conjecture, we used the {\textsc{screens}} package \citep{marten_h_van_kerkwijk_2022_7455536} to simulate a one-dimensional phase screen where rays scattered at positive angles were 30\% brighter than those scattered at negative angles for a pulsar with the same astrometrical properties as PSR J1136+1551 under the assumption that the pulsar motion dominates the effective velocity and using the same observing setup as our observations. We then repeated the simulation at 50 fractional screen distances between $0.5-0.8$ while using the same image on the sky in each setup. This fractional screen distance range was chosen to allow for significant exploration of the phenomenon along the LOS while limiting the proximity of the screen to Earth so that we could safely ignore the motion of the Earth in our simulation. 
\begin{figure}[!ht]
    \centering
    \includegraphics[width=0.5\textwidth]{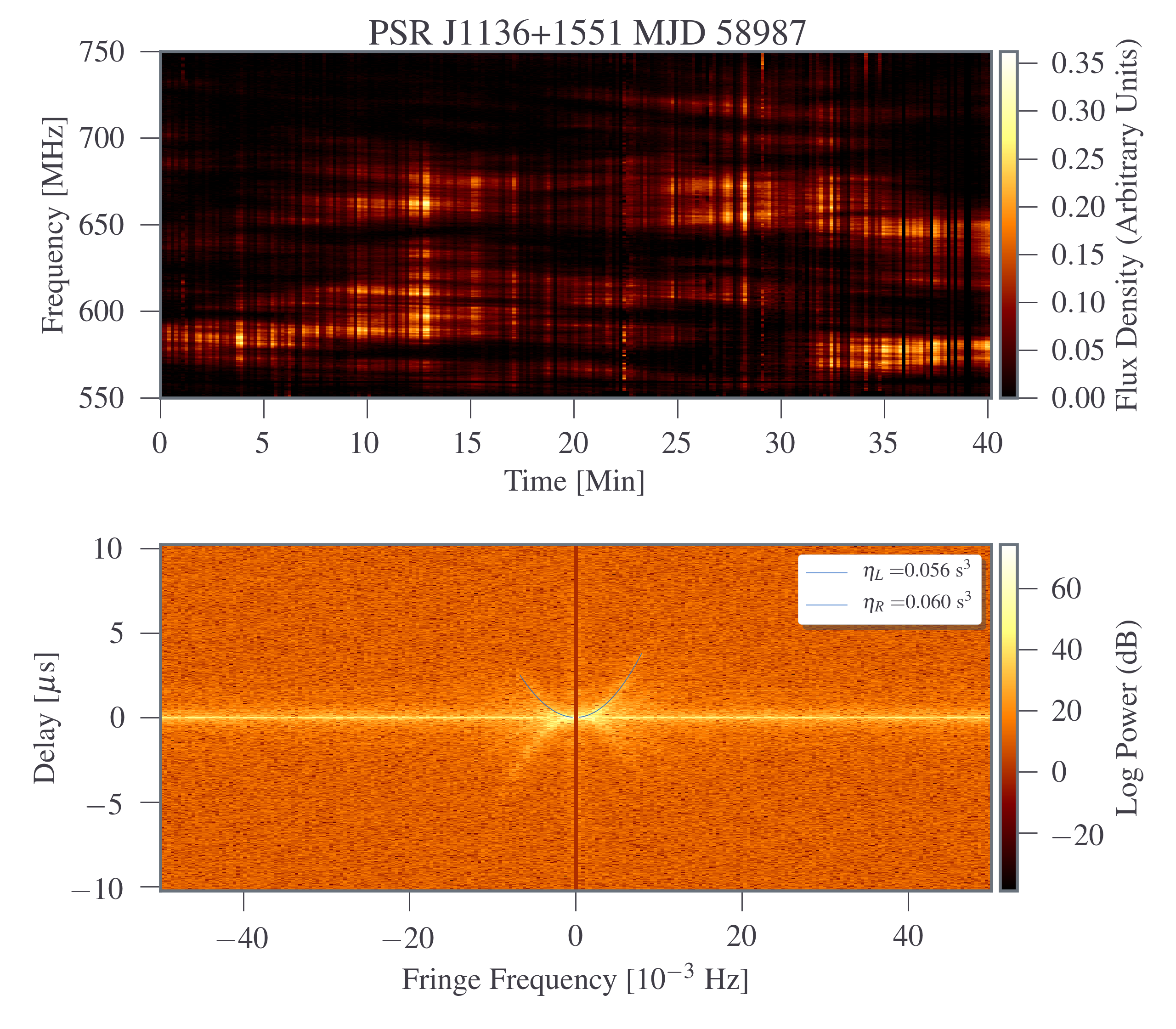} %
    \caption{Dynamic (top) and secondary (bottom) spectra of PSR J1136+1551 centered at 650 MHz on MJD 58987. The top half of the secondary spectrum shows the overlaid arc fits in blue. \iffalse Scaled uncertainties of the arc curvature can be found in Table \ref{gen_results}.\fi}
    \label{ex_epoch_1}
\end{figure}
\par
\begin{figure}[!ht]
    \centering
    \includegraphics[width=0.5\textwidth]{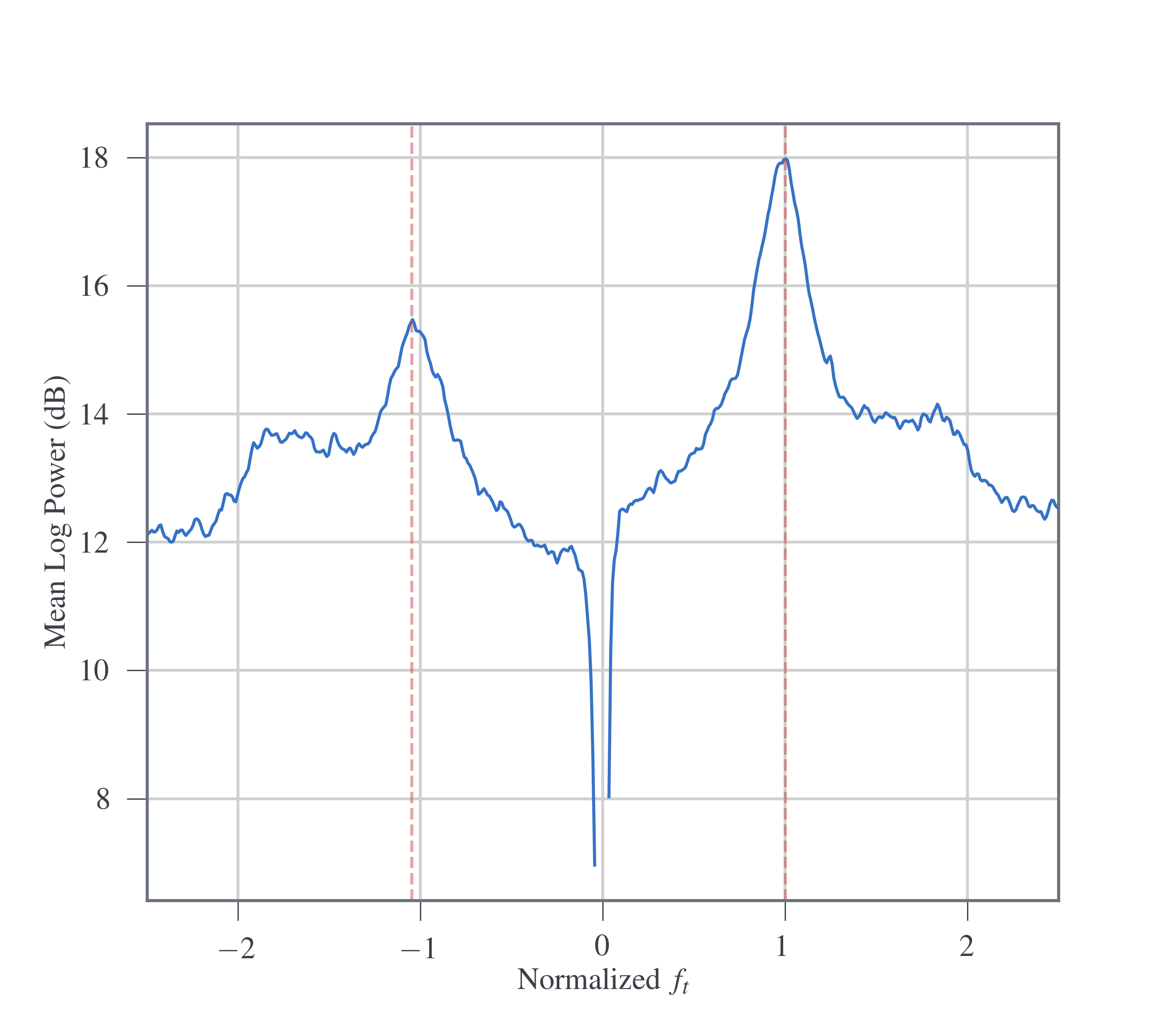}%
    \caption{Normalized secondary spectrum power profile of PSR J1136+1551 centered at 650 MHz on MJD 58987. The vertical dashed lines indicate where the arcs fall on the normalized delay axis.}%
    \label{ex_epoch_2}
\end{figure}

\begin{figure}[!ht]
    \centering
    \includegraphics[scale = 0.59]{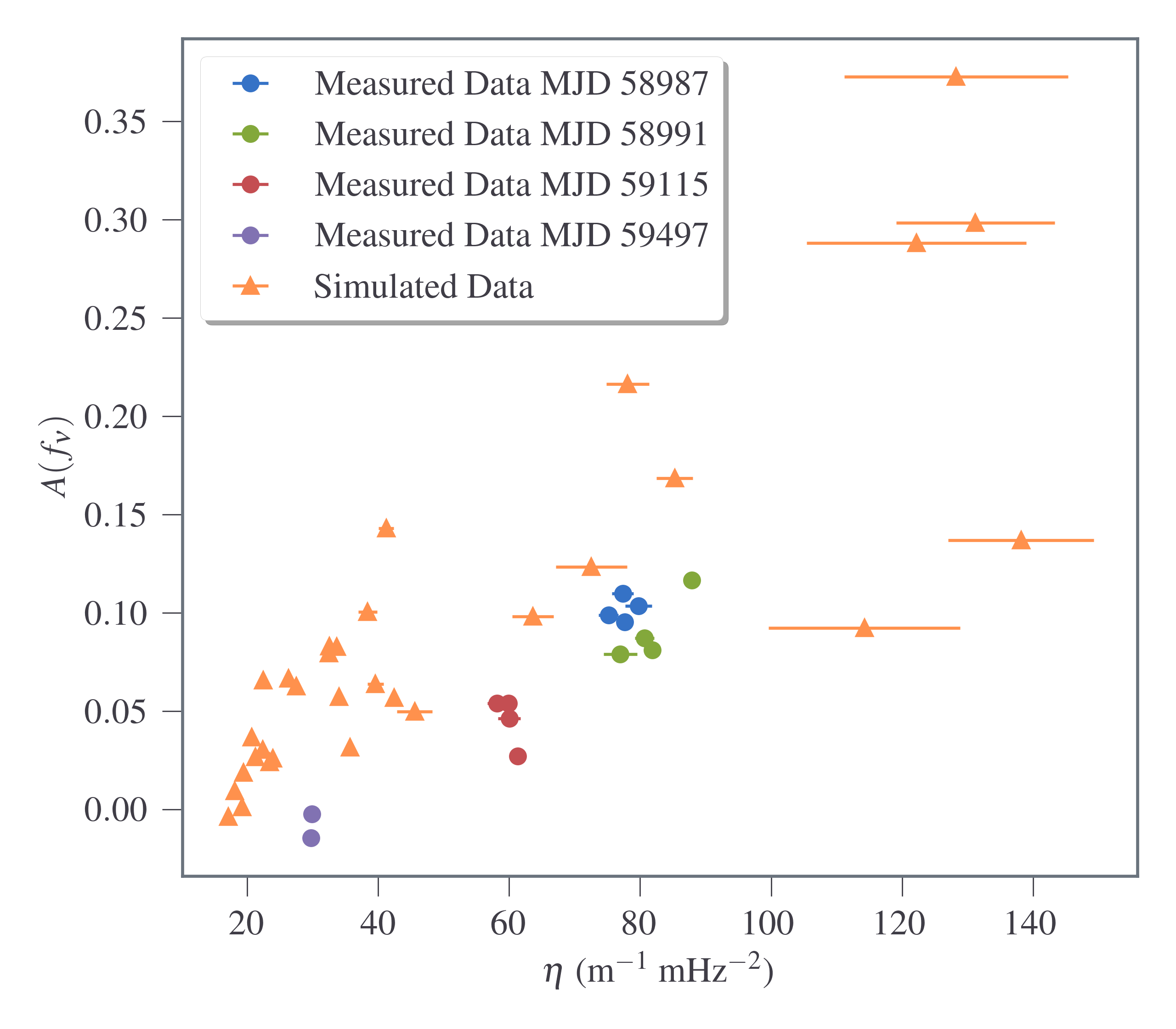}
    \caption{Scatter plot showing simulated (orange triangles) and measured (dots of various colors) arc curvatures and the corresponding asymmetry indices for all measurements of PSR J1136+1551 taken with Band 4. The three distinct clumps in the measured data are the result of the observations being dominated by three different scattering screens.}
    \label{scatter_cor}
\end{figure}

 An example dynamic and secondary spectrum pair from the observation on MJD 58987 is shown in Figure \ref{ex_epoch_1}, with its corresponding normalized secondary spectrum power profile, which is used to determine the asymmetry index, shown in Figure \ref{ex_epoch_2}, while the scatter plot showing the relation between measured arc curvature and arc asymmetry index across all measurements taken in the 650 MHz band, along with our simulated data, is shown in Figure \ref{scatter_cor}. As seen in Figure \ref{scatter_cor}, despite the image on the sky remaining the same in each of our simulations, the asymmetry gets progressively larger as we go to higher curvatures. For this reason we conclude this correlation to be the result of a bias in the Hough transform approach for measuring arc curvatures and the resulting normalized secondary spectra. Of particular note in Figure \ref{scatter_cor} are the three distinct clumps in our measured data, which we believe are the result of our observations being dominated by a different scattering screen at each epoch (two of our observations were taken four days apart, and so are dominated by the same screen). It is likely that this pulsar's at least six known scattering screens are the main reason why we were able to see this correlation in our data in the first place, as individual scattering screens likely do not vary enough in distance over time for this trend to become apparent. Indeed, the limited number of pulsars with multiple known screens is probably the main reason why this trend has not been reported in earlier studies.
\subsection{J1509+5531}
\begin{figure}[!ht]
    \centering
    \includegraphics[scale = 0.55]{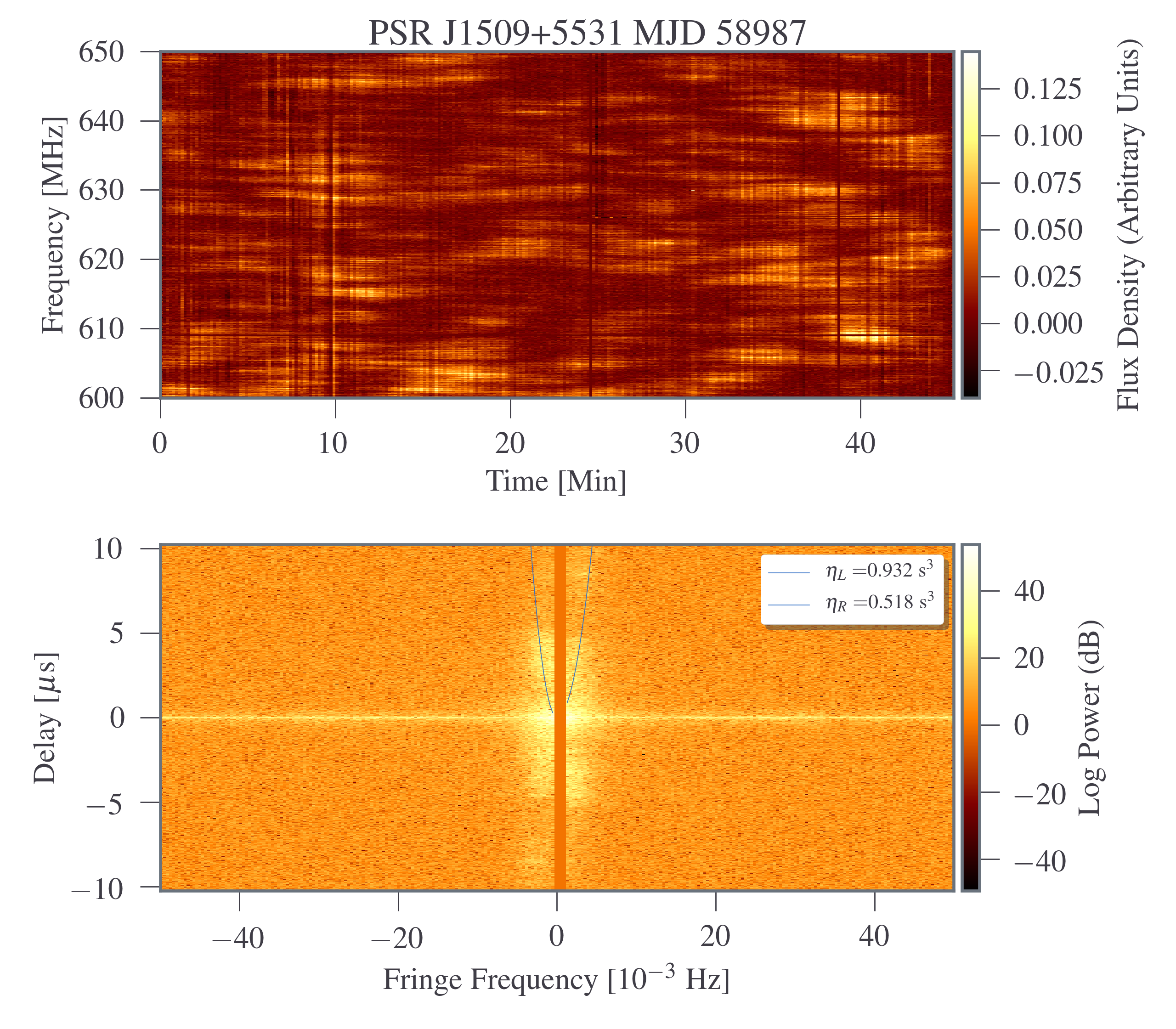}
    \caption{Dynamic (top) and secondary (bottom) spectra of PSR J1509+5531 centered at 625 MHz on MJD 58987. The top half of the secondary spectrum shows the overlaid arc fits in blue. \iffalse Scaled uncertainties of the arc curvature can be found in Table \ref{gen_results}.\fi During this period of observations, visible arcs were considerably narrower than later observations.}
    \label{1509_fig}
\end{figure}

\begin{figure}[!ht]
    \centering
    \includegraphics[scale = 0.55]{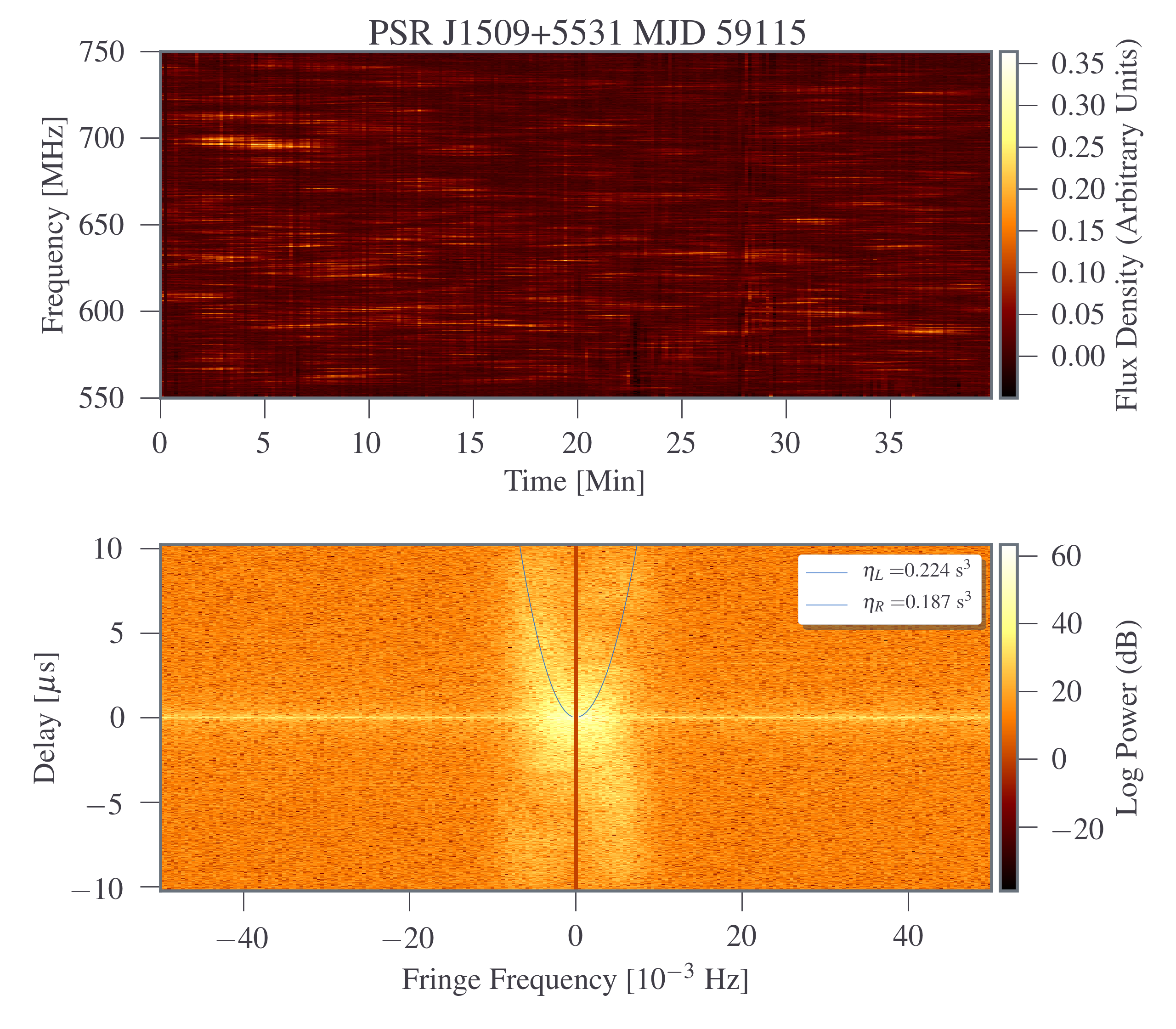}
    \caption{Dynamic (top) and secondary (bottom) spectra of PSR J1509+5531 centered at 650 MHz on MJD 59115. The top half of the secondary spectrum shows the overlaid arc fits in blue. \iffalse Scaled uncertainties of the arc curvature can be found in Table \ref{gen_results}.\fi During this period of observations, visible arcs were considerably wider than earlier observations.}
    \label{1509_fig_2}
\end{figure}

In the observations of this pulsar in the 650 MHz band, all secondary spectra featured patchy rather than continuous arcs, particularly in the left arm. This patchiness indicates a detection of this pulsar's arclets, which result from substructures in the ISM thought to arise from scattering interference between an inhomogeneously scattered distribution of material and some distinct offset region \citep{walker_stinebring,cordes_2006_refraction}. Under the assumption that these substructures are lens-like, they are expected to be roughly AU in scale \citep{hsa+05}. Unique to these arclets is their distinctly flat nature, which has been attributed to its exceptionally high transverse velocity of 960$^{+61}_{-64}$ km s$^{-1}$ \citep{Chatterjee_2009}. Interestingly, the arc curvatures measured in the last two epochs (MJDs 59115 and 59497) are a factor of two to three times smaller than the first two epochs (MJDs 59064 and 58987), possibly indicating a detection of multiple scattering screens along the LOS to this pulsar. This result augments the results of \cite{mult_screen_1508}, who also found significant variability along the LOS to this pulsar during the same period of time and propose a double screen model as a possible explanation. An example observation from the earlier two epochs is shown in Figure \ref{1509_fig}, while an example from the later two epochs is shown in Figure \ref{1509_fig_2}.

\subsection{J1645-0317}

\begin{figure}
    \centering
    \includegraphics[scale=0.58]{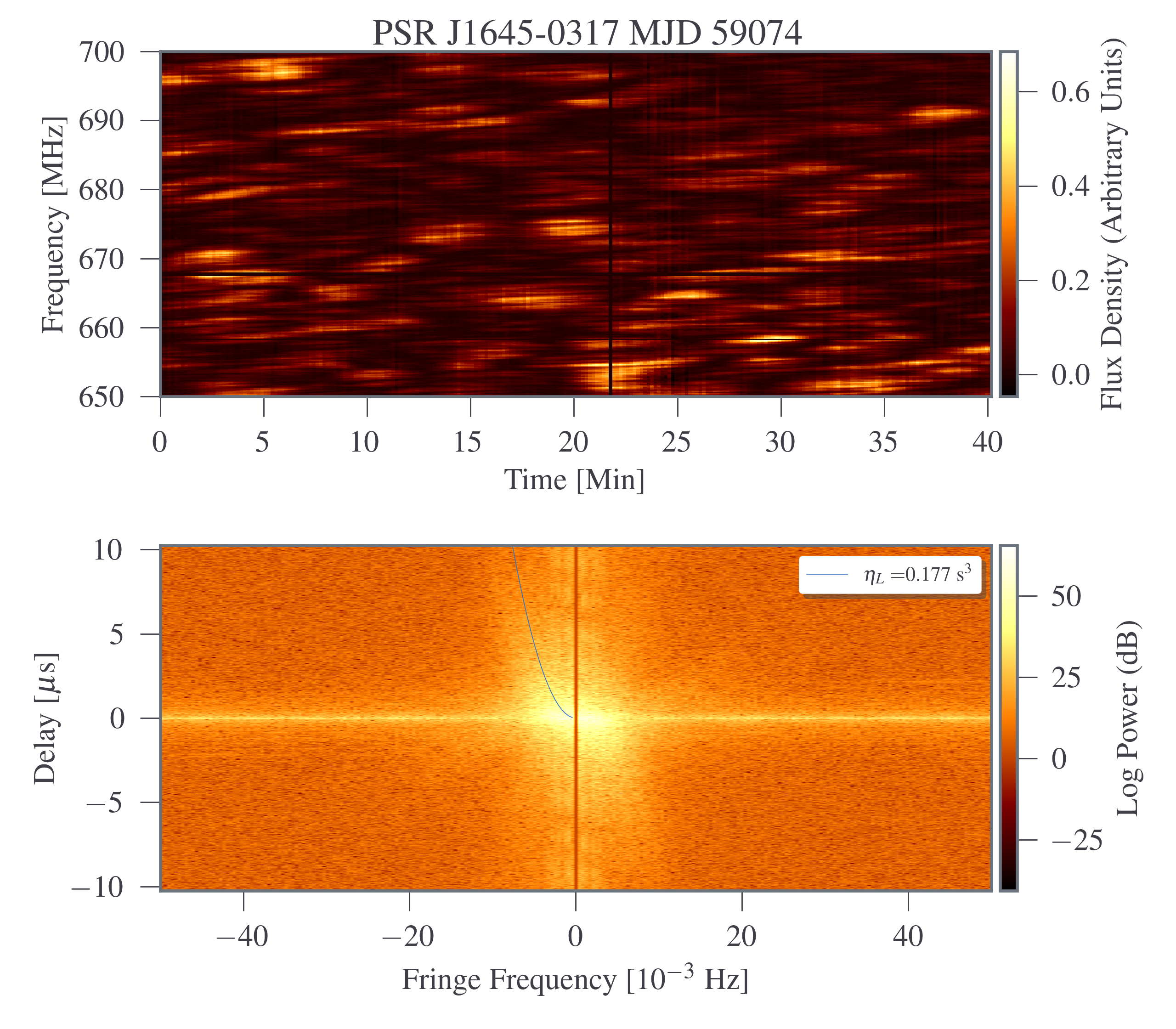}
    \caption[Example dynamic and secondary spectrum for PSR J1645-0317]{Dynamic (top) and secondary (bottom) spectra of PSR J1645-0317 centered at 675 MHz on MJD 59074. The top half of the secondary spectrum shows the overlaid arc fits in blue. \iffalse Scaled uncertainties of the arc curvature can be found in Table \ref{gen_results}.\fi}
    \label{1645_fig}
\end{figure}

In the one epoch for which we were able to resolve a scintillation arc in this pulsar, its power was found to be highly concentrated towards the origin, indicating strong scintillation with the majority of the power in its brightness distribution originating from around $\theta=0$ \citep{cordes_2006_refraction}. There is also considerably more power on the left side of the fringe frequency axis, indicating strong refraction occurring during this epoch. An example observation is shown in Figure \ref{1645_fig}.

\subsection{J1932+1059}
Due to having the lowest DM in our survey, this pulsar showed the least variation in arc curvature from epoch to epoch across all frequencies. Its close proximity to Earth also resulted in wide scintles in frequency, leading to high scintle resolution and very bright, narrow, and well defined arcs. The sharpness of these arcs may also indicate scattering that is highly anisotropic along the LOS \citep{walker_2004,cordes_2006_refraction}, as well as originating from a discrete, localized source \citep{OG_arcs}. Overall this was our most consistent pulsar in all aspects of scintillation. 
\par
\begin{figure}
    \centering
    \includegraphics[scale=0.58]{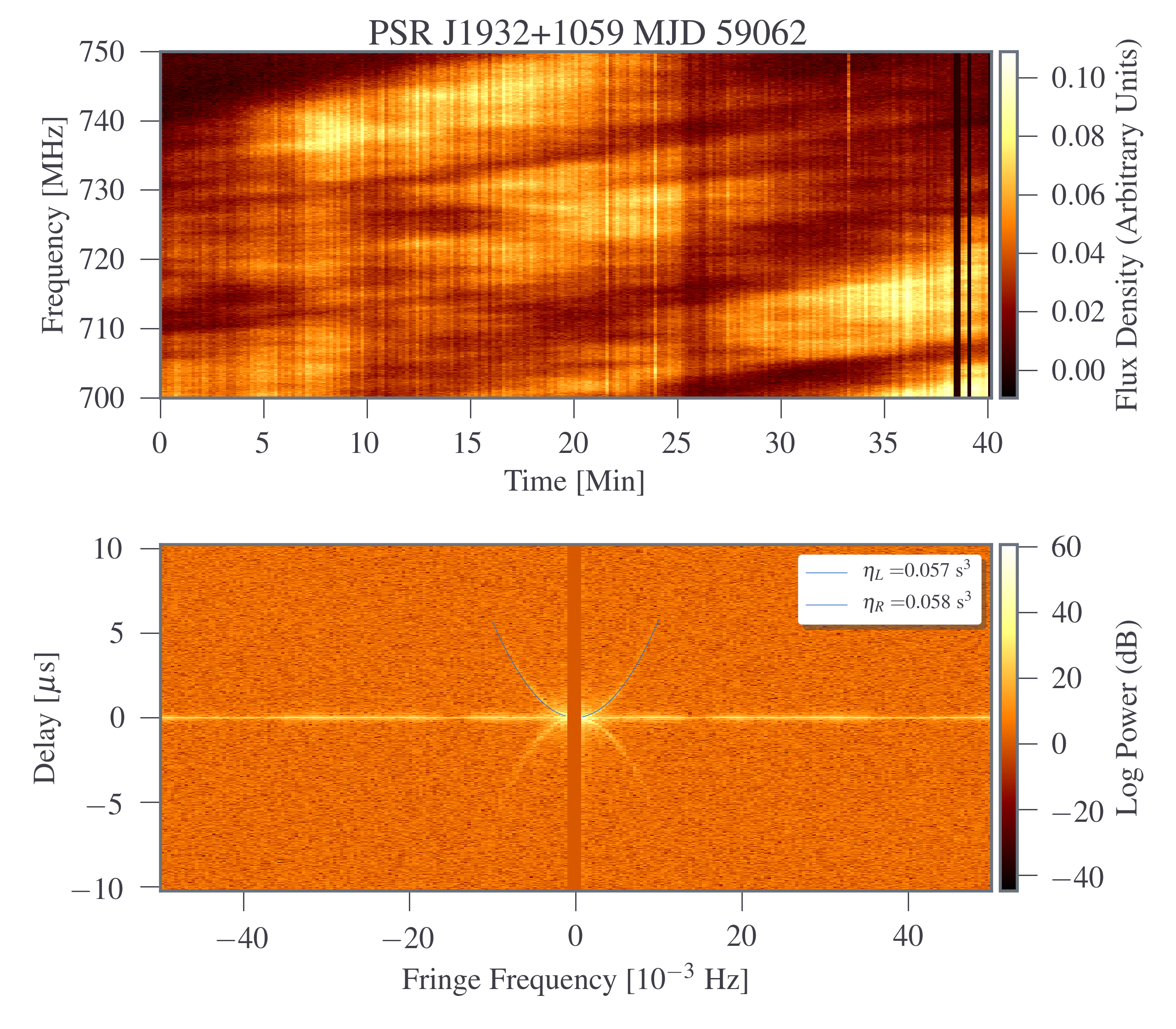}
    \caption{Dynamic (top) and secondary (bottom) spectra of PSR J1932+1059 centered at 725 MHz on MJD 58987. The top half of the secondary spectrum shows the overlaid arc fits in blue. \iffalse Scaled uncertainties of the arc curvature can be found in Table \ref{gen_results}.\fi}
    \label{1932_fig}
\end{figure}

 This consistency lines up with its other astrophysical parameters, as its dispersion measure of 3.18 pc cm$^{-3}$ \citep{atnf_7,atnf_3} was  the lowest in our survey and its transverse velocity of 152 km s$^{-1}$ \citep{atnf_1,atnf_3} was the second lowest. While its transverse velocity is a bit larger than that of PSR J0630$-$2834 and their distances are almost equivalent, PSR J0630$-$2834 has a dispersion measure 10 times higher than PSR J1932{+}1059 \citep{atnf_7,atnf_6,atnf_3}. This means that a much denser ISM was sampled in PSR J0630$-$2834 than in PSR J1932{+}1059, meaning that PSR J1932{+}1059 had decisively the least amount of ISM sampled over our survey, making it the least likely to experience large scintillation-related variations. An example observation is shown in Figure \ref{1932_fig}.

\par Like in the case of PSR J1136+1551, two of our three Gaussian frequency ACF-derived scintillation bandwidth scaling indices are shallower than the measured index using lower observing frequencies from \cite{wu_2022}. In the case of this pulsar, our observations also overlap theirs in time, meaning we would expect similar LOS behavior, although it has been observed that the measured scaling index along the LOS to a given pulsar can vary over time \citep{Levin_Scat,turner_scat}. Interestingly, when comparing the discrepancies between our scaling indices and \cite{wu_2022} using the time of year in which the observations were made, the closer our observing epochs are to theirs as a fraction of a year, the more our measured scaling indices agree. More specifically, looking at the difference between epochs in days relative to a calendar year, they measured a scaling index of $4.00\pm 0.37$, and we found scaling indices of $1.89 \pm 0.46$, $2.02 \pm 0.42$, and $3.71 \pm 0.37$ with day differences of 151, 149, and 82 from their epoch, respectively. It is known that other ISM-related quantities such arc curvature \citep{Main, McKee_2022} and dispersion measure \citep{Hazboun_2022} have annual variations, and it is expected that scintillation bandwidth should correlate with dispersion measure \citep{event, Coles_2015,lentati,mckee}, meaning we might expect some annual trend in scintillation bandwidth measurements as well. However, it seems unlikely that the annual variations in electron density contributions from the solar wind would noticeably affect measured ISM turbulence, much less to the degree seen in our data.

\subsection{J2048-1616}
\begin{figure}
    \centering
    \includegraphics[scale=0.58]{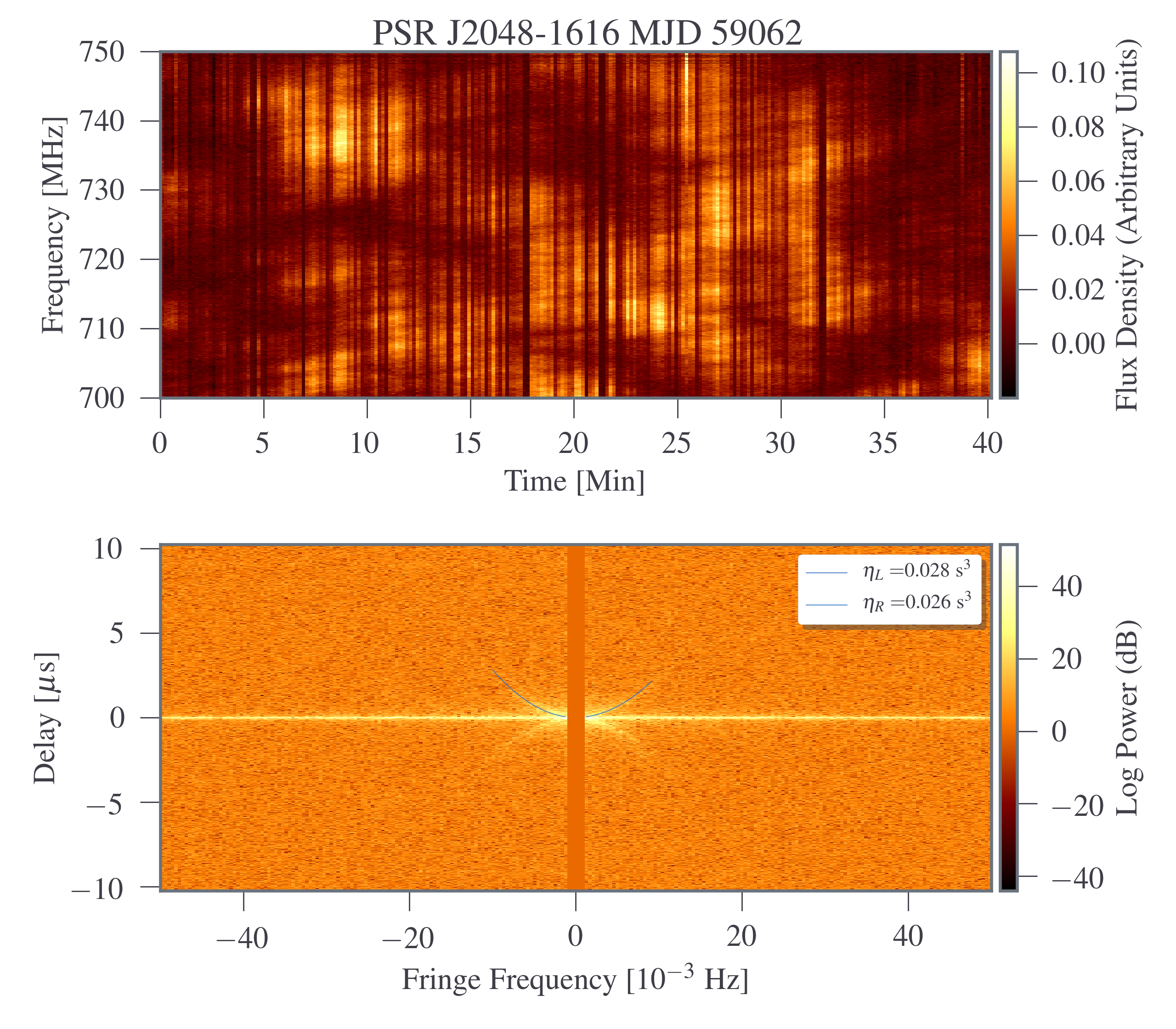}
    \caption{Dynamic (top) and secondary (bottom) spectra of PSR J2048-1616 centered at 725 MHz on MJD 58987. The top half of the secondary spectrum shows the overlaid arc fits in blue.}
    \label{2048_fig}
\end{figure}

This pulsar exhibited strong pulse intensity variability at all observing frequencies, resulting in significant flux density variation within individual scintles. The scintillation arcs for this pulsar exhibit high levels of asymmetry, potentially a consequence of refraction. This strong asymmetry is also accompanied by a highly asymmetric scaling of the arc curvature in both arms (see Table \ref{scaling_results}), with the left and right arms exhibiting noticeably steeper and shallower arc curvature scaling indices, respectively, than would be expected as a consequence of pure plasma refraction. This asymmetry diminishes noticeably with observing frequency, visible both by eye and by looking at $A(f_{\nu})$ as a function of frequency, possibly a consequence of weaker refraction at higher observing frequencies. Additionally, the thickness of these arcs also sharply decreases with frequency, being barely visible over the power near the origin at low frequencies and very sharply defined at higher frequencies. \cite{stine_ock} examined how the width of scintillation arcs change with frequency in PSR J1136+1151, also observing a decrease in thickness with frequency. They concluded these effects could be explained by a linear, one-dimensional brightness function and an anisotropic brightness distribution, which describe how the brightness of the scattered image changes with the angular deflection of the scattered ray in 1D and 2D, respectively. An example observation is shown in Figure \ref{2048_fig}.

\section{Conclusions \& Future Work}
\label{sec:conclusion}
We performed simultaneous dual-frequency observations of six bright canonical pulsars using the uGMRT. We extracted scintillation arc, bandwidth, and timescale measurements for each of these pulsars to examine a variety of science. We examined how arc curvature scaled with frequency and found our observations to be consistent with the index predicted by theory and demonstrated in the literature, while at the same time using a more astronomically ideal setup to perform these measurements. We also measured scintillation bandwidth and scintillation timescale scaling indices for five of our six pulsars and found indices consistent with or shallower than Kolmogorov turbulence, agreeing with previous literature. We examine the relation between arc curvature and scintillation bandwidth and timescale within epochs and find strong correlation between the measured curvature and pseudo-curvature, indicating our assumptions of one-dimensional, thin screen scattering are justified. Finally, we find an interesting and strong correlation between arc curvature and arc asymmetry in PSR J1136+1551, which we attribute to a bias in the Hough transform method for measuring arc curvatures.
\par This study demonstrates the value of array-based telescopes such as uGMRT to the pulsar astronomy community, as well as the strengths of simultaneous multiband studies of pulsars and the wide variety of science that can be done with such an approach. This also shows strong promise for the future observations using ultrawideband (UWB) receivers, which are coming online at instruments such as the Green Bank Telescope.
\par We thank the staff at the uGMRT who have made these observations possible. The uGMRT is run by the National Centre for Radio Astrophysics of the Tata Institute of Fundamental Research. We would also like to thank Daniel Baker, Marten van Kerkwijk, Ue-Li Pen, and the rest of the scintillometry group at University of Toronto for crucial discussions surrounding our scintillation arc asymmetry analyses. We gratefully acknowledge support of this effort from the NSF Physics Frontiers Center grants 1430284 and 2020265 to NANOGrav. Some of the data processing in this work utilized the resources of the Bowser computing cluster at West Virginia University.

\par \textit{Software}: \textsc{scintools} \citep{scintools}, \textsc{screens} \citep{marten_h_van_kerkwijk_2022_7455536}, \textsc{pypulse} \citep{pypulse}, \textsc{scipy} \citep{scipy}, \textsc{numpy} \citep{numpy}, and \textsc{matplotlib} \cite{matplotlib}.
\bibliography{gmrt_arc_paper.bib}{}
\bibliographystyle{aasjournal}
\end{document}